      \documentstyle[12pt, righttag]
{amsart}

    \input amssym.def
    \input amssym

    \newtheorem{propo}{Proposition}
    \newtheorem{theo}[propo]{Theorem}
    \newtheorem{lemma}[propo]{Lemma}
    \newtheorem{corol}[propo]{Corollary}

    \theoremstyle{definition}
    \newtheorem{defi}[propo]{Definition}

    \theoremstyle{remark}
    \newtheorem{rema}[propo]{Remark}

\numberwithin{equation}{section}
\numberwithin{propo}{section}

\begin{document}
    \title[A nonmeromorphic extension of the moonshine module]{A
nonmeromorphic extension of the moonshine module
vertex operator algebra}
    \author{Yi-Zhi Huang}
    \address{Department of Mathematics, University of
Pennsylvania, Philadelphia, PA 19104
and
Department of Mathematics\\ Rutgers University\\
New Brunswick, NJ 08903 (current address)}
    \email{yzhuang@@math.rutgers.edu}
    \thanks{This research is supported in part by NSF grant
DMS-9301020 and by DIMACS, an NSF Science and Technology Center funded
under contract STC-88-09648.}
	\bibliographystyle{alpha}
	\maketitle

	\newcommand{\nno}{\nonumber}
	\newcommand{\lbar}{\bigg\vert}
	\newcommand{\p}{\partial}
	\newcommand{\dps}{\displaystyle}
	\newcommand{\bra}{\langle}
	\newcommand{\ket}{\rangle}
 \newcommand{\res}{\mbox{\normalshape Res}}
 \newcommand{\epf}{\hspace{2em}$\Box$}
 \newcommand{\epfv}{\hspace{1em}$\Box$\vspace{1em}}
\newcommand{\nord}{\mbox{\scriptsize ${\circ\atop\circ}$}}
\newcommand{\wt}{\mbox{\normalshape wt}\ }

 \makeatletter
\newlength{\@pxlwd} \newlength{\@rulewd} \newlength{\@pxlht}
\catcode`.=\active \catcode`B=\active \catcode`:=\active \catcode`|=\active
\def\sprite#1(#2,#3)[#4,#5]{
   \edef\@sprbox{\expandafter\@cdr\string#1\@nil @box}
   \expandafter\newsavebox\csname\@sprbox\endcsname
   \edef#1{\expandafter\usebox\csname\@sprbox\endcsname}
   \expandafter\setbox\csname\@sprbox\endcsname =\hbox\bgroup
   \vbox\bgroup
  \catcode`.=\active\catcode`B=\active\catcode`:=\active\catcode`|=\active
      \@pxlwd=#4 \divide\@pxlwd by #3 \@rulewd=\@pxlwd
      \@pxlht=#5 \divide\@pxlht by #2
      \def .{\hskip \@pxlwd \ignorespaces}
      \def B{\@ifnextchar B{\advance\@rulewd by \@pxlwd}{\vrule
         height \@pxlht width \@rulewd depth 0 pt \@rulewd=\@pxlwd}}
      \def :{\hbox\bgroup\vrule height \@pxlht width 0pt depth
0pt\ignorespaces}
      \def |{\vrule height \@pxlht width 0pt depth 0pt\egroup
         \prevdepth= -1000 pt}
   }
\def\endsprite{\egroup\egroup}
\catcode`.=12 \catcode`B=11 \catcode`:=12 \catcode`|=12\relax
\makeatother

\def\hboxtr{\FormOfHboxtr} 
\sprite{\FormOfHboxtr}(25,25)[0.5 em, 1.2 ex] 

:BBBBBBBBBBBBBBBBBBBBBBBBB |
:BB......................B |
:B.B.....................B |
:B..B....................B |
:B...B...................B |
:B....B..................B |
:B.....B.................B |
:B......B................B |
:B.......B...............B |
:B........B..............B |
:B.........B.............B |
:B..........B............B |
:B...........B...........B |
:B............B..........B |
:B.............B.........B |
:B..............B........B |
:B...............B.......B |
:B................B......B |
:B.................B.....B |
:B..................B....B |
:B...................B...B |
:B....................B..B |
:B.....................B.B |
:B......................BB |
:BBBBBBBBBBBBBBBBBBBBBBBBB |

\endsprite

\def\shboxtr{\FormOfShboxtr} 
\sprite{\FormOfShboxtr}(25,25)[0.3 em, 0.72 ex] 

:BBBBBBBBBBBBBBBBBBBBBBBBB |
:BB......................B |
:B.B.....................B |
:B..B....................B |
:B...B...................B |
:B....B..................B |
:B.....B.................B |
:B......B................B |
:B.......B...............B |
:B........B..............B |
:B.........B.............B |
:B..........B............B |
:B...........B...........B |
:B............B..........B |
:B.............B.........B |
:B..............B........B |
:B...............B.......B |
:B................B......B |
:B.................B.....B |
:B..................B....B |
:B...................B...B |
:B....................B..B |
:B.....................B.B |
:B......................BB |
:BBBBBBBBBBBBBBBBBBBBBBBBB |

\endsprite

\begin{abstract}
We describe a natural structure of an abelian intertwining algebra (in
the sense of Dong and Lepowsky) on the direct sum of the untwisted
vertex operator algebra constructed {}from the Leech lattice and its
(unique) irreducible twisted module.  When restricting ourselves to
the moonshine module, we obtain a new and conceptual proof that the
moonshine module has a natural structure of a vertex operator algebra.
This abelian intertwining algebra also contains an irreducible twisted
module for the moonshine module with respect to the obvious
involution. In addition, it contains a vertex operator superalgebra
and a twisted module for this vertex operator superalgebra with
respect to the involution which is the identity on the even subspace
and is $-1$ on the odd subspace. It also gives the superconformal
structures observed by Dixon, Ginsparg and Harvey.
\end{abstract}

The relation between the modular function $J(q)=j(q)-744$ and
dimensions of certain representations of the Monster was first noticed
by McKay and Thompson (see \cite{Th}). Based on these
observations, McKay and Thompson conjectured the existence of a
natural ($\Bbb{Z}$-graded) infinite-dimensional representation of the
Monster group such that its graded dimension as a $\Bbb{Z}$-graded
vector space is equal to $J(q)$. In the famous paper \cite{CN} by
Conway and Norton, remarkable numerology between McKay-Thompson series
(graded traces of elements of the Monster acting on the conjectured
infinite-dimensional representation of the Monster) and modular
functions was collected, and surprising conjectures about those
modular functions occured were presented. The Monster was
constructed by Griess \cite{G} later but the mysterious connection
between the Monster and modular functions was still not expained.
In \cite{FLM1}, Frenkel, Lepowsky and Meurman
constructed a natural infinite-dimensional representation of the
Monster (called the moonshine module and denoted $V^{\natural}$) using
the method of vertex operators. The moonshine module constructed by
them provided a remarkable conceptual framework towards the
understanding  of monstrous moonshine. In particular,
some of the numerology and conjectures in
\cite{CN} were expained and proved in \cite{FLM1}.
Motivated partly by \cite{FLM1}, Borcherds \cite{B1} developed a
general theory of vertex operators based on an even lattice. {}From
this general theory, he axiomized the notion of ``vertex algebra'' and
using the results announced in \cite{FLM1}, he stated that the
moonshine module $V^{\natural}$ has a structure of such an algebra.
In \cite{FLM2}, Frenkel, Lepowsky and Meurman
proved that the moonshine module $V^{\natural}$ has a natural
structure of vertex operator algebra and the Monster is the
automorphism group of this vertex operator algebra.  Their proof is
very involved and uses triality and some results in group theory.
On the other hand, $V^{\natural}$ can be viewed as a substructure of a
$\Bbb{Z}_{2}$-orbifold conformal field theory. Using techniques
developed in string theory, Dolan, Goddard and Montague \cite{DGM1}
gave another proof that $V^{\natural}$ has a natural structure of
vertex operator algebra.  Their proof works for a class of
$\Bbb{Z}_{2}$-orbifold theories and thus allows them to give a further
interpretation of Frenkel-Lepowsky-Meurman's triality \cite{DGM2}
\cite{DGM3} (see also \cite{L}).
But their proof is still very technical and complicated.

Using the moonshine module $V^{\natural}$ constructed by Frenkel,
Lepowsky and Meurman, the no-ghost theorem in string theory and the
theory of generalized Kac-Moody algebras (or Borcherds algebras),
Borcherds \cite{B2} completed the proof of the monstrous moonshine
conjecture in \cite{CN}.  Part of Borcherds' proof has been simplified
recently by Jurisich \cite{J} and by Jurisich, Lepowsky and Wilson
\cite{JLW}.  But the last step of Borcherds' proof uses some case by case
identification which is conceptually unsatisfactory. Also it seems
that the methods used in the proof cannot be used to prove the
generalized moonshine conjecture \cite{N}. In \cite{Tu1}, \cite{Tu2}
and
\cite{Tu3}, Tuite showed that the monstrous moonshine conjecture, and
the generalized moonshine conjecture in some special cases, can
be understood by using Frenkel-Lepowsky-Meurman's uniqueness
conjecture on $V^{\natural}$ and some conjectures in the
mathematically yet-to-be-established orbifold conformal field theory.
In particular, he pointed out that nonmeromorphic operator algebras
for orbifold theories of the Leech lattice theory and the moonshine
module, which are the foundation of his idea, are still to be
constructed, even in the original simplest $\Bbb{Z}_{2}$-orbifold
case. The importance of these nonmeromorphic operator algebras is that
they give the whole genus-zero chiral parts of the orbifold conformal
theories.  Also the automorphism groups of these nonmeromorphic
operator algebras might be of interest.

In this paper, we describe a construction of the nonmeromorphic
extension of $V^{\natural}$ in the original simplest
$\Bbb{Z}_{2}$-orbifold case.  The main tools which we use are the
decompositions of the untwisted vertex operator algebra associated to
the Leech lattice and its irreducible twisted module into direct sums
of irreducible modules for a tensor product of the Virasoro vertex
operator algebra with central charge $\frac{1}{2}$ obtained by Dong,
Mason and Zhu (\cite{DMZ},
\cite{D3}),  and
the theory of tensor products of modules for a vertex operator algebra
developed by Lepowsky and the author (\cite{HL1}--\cite{HL5},
\cite{H2}).  Precisely speaking, we describe a natural structure of
an abelian intertwining algebra
(in the sense of Dong and Lepowsky \cite{DL}) on the direct sum of
the untwisted vertex operator algebra constructed {}from the Leech
lattice and its (unique) irreducible twisted module with respect to an
involution induced {}from a reflection of the Leech lattice
(see Theorem \ref{main}).
Our construction and proof are conceptual, that is, every step is
natural in the theory of vertex operator algebras. In particular, when
restricting ourselves to the moonshine module, we obtain a new and
conceptual proof that the moonshine module has a natural structure of
vertex operator algebra.
This abelian
intertwining algebra also contains a (unique) irreducible
twisted module (which has also been obtained by Dong and Mason using a
different method) for the moonshine module with respect to the obvious
involution, a vertex operator superalgebra and a twisted module for
this vertex operator superalgebra with respect to the  involution
which is the identity on the even subspace and is $-1$ on
the odd subspace.
This abelian intertwining algebra also gives the superconformal
structures observed by Dixon, Ginsparg and Harvey \cite{DGH}.  We
define a superconformal vertex operator algebra to be an abelian intertwining
 algebra of a certain type equipped with an element
which together with the Virasoro element generates a super-Virasoro
algebra. Then the abelian intertwining algebra structure above
together with any one of the superconformal structures of Dixon,
Ginsparg and Harvey is a superconformal vertex operator algebra.

Note that the tensor product theory can be applied to modules for any rational
vertex operator algebra satisfying certain conditions described in
\cite{H2} and
\cite{HL4} (see also Subsection 2.2 below). In the present paper,
the results in \cite{DMZ} and \cite{D3} are used to show that the
tensor product theory can be used and to calculate the fusion rules.
Thus for other orbifold theories and conformal field theories, if the
conditions to use the tensor product theory are satisfied and the
fusion rules can be calculated, we can also construct the
nonmeromorphic operator algebra in the same way.  In particular, it
might be possible to prove the result of \cite{DM} using the tensor
product theory.

Since the present paper uses almost all basic concepts and results in
the algebraic theory of vertex operator algebras, it is impossible to
give a complete exposition to those basic materials which we need.
Therefore we assume that the reader is familiar with the basic notions
and results in the theory of vertex operator algebras. For details,
see \cite{FLM1} and \cite{FHL}.  For more material on the axiomatic
aspects of the theory of vertex operator algebras, we shall refer the
reader to the appropriate references. We shall, however, review
briefly the results on the moonshine module and other related results
which we need. We shall also give a brief account of the part which we
need of the tensor product theory of modules for a vertex operator
algebra.

This paper is organized as follows: Section 1 is a review of the
constructions of and results on the moonshine module and related
structures. Section 2 is a review of a certain part of the tensor
product theory for modules for a vertex operator algebra. The
construction of the nonmeromorphic extension is sketched in Section 3.
Some consequences of this nonmeromorphic extension, including the
vertex operator algebra structure on the moonshine module, the twisted
module for the moonshine module, a vertex operator superalgebra and its
twisted module in this extension, and the superconformal
structures of Dixon, Ginsparg and Harvey, are described in Section 4.

The details of the proofs of the results described in this paper
will be published elsewhere.

\subsection*{Acknowledgement} I would like to thank Chongying Dong
 for discussions and Jim Lepowsky for discussions, comments and especially
explanations of results in \cite{FLM1} and \cite{FLM2}.
I would also like to thank Wanglai Li for
pointing out a mistake in the description of the nonmeromorphic extension
in an earlier version of this paper.

\section{Leech lattice theory and the moonshine module}

In this section we review briefly the constructions of the untwisted
vertex operator algebra, associated to the Leech lattice, its
irreducible twisted module and the moonshine module $V^{\natural}$. We
also review some results on these algebras and modules and related
structures. For more details, the reader is referred to \cite{FLM2},
\cite{DMZ}, \cite{D1},  \cite{D2}, \cite{D3}.

\subsection{The Golay code $\cal{C}$}
Let $\Omega=\{S_{1}, \dots, S_{n}\}$ be a finite set. A {\it (binary
linear) code} on $\Omega$ is a subspace of the vector space ${\cal
P}(\Omega)=\coprod_{i=1}^{n}\Bbb{Z}_{2}S_{i}$ over $\Bbb{Z}_{2}$
spanned by elements of $\Omega$. Any element $S$ of $\cal{P}(\Omega)$
is a linear combination of $S_{1},
\dots, S_{n}$.  The number of nonzero coefficients is called the {\it
weight} of $S$ and is denoted as $|S|$. A code $\cal{S}$ is said to
be of {\it type} II if $n\in 4\Bbb{Z}$, $|S|\in 4\Bbb{Z}$ for all
$S\in \cal{S}$ and $S_{1}+\cdots +S_{n}\in \cal{S}$. The usual dot
product for a vector space with a basis gives a natural nonsingular
symmetric bilinear form on $\cal{P}(\Omega)$.  The annihilator of a
code $\cal{S}$ in $\cal{P}(\Omega)$ with respect to this bilinear
form is again a code. It is called the {\it dual code} of $\cal{S}$
and is denoted $\cal{S}^{\circ}$. A code is called {\it self-dual} if
it is equal to its dual code.

\begin{theo}
There is a self-dual code of type II on a $24$-element set such that
it has no elements of weight $4$. It is unique up to isomorphism.
\end{theo}

The code in this theorem is called the {\it Golay code} and is denoted
$\cal{C}$.

\subsection{The Leech lattice $\Lambda$}
A {\it (rational) lattice of rank} $n\in \Bbb{N}$ is a rank $n$ free
abelian group $L$ equipped with a rational-valued symmetric
$\Bbb{Z}$-bilinear form $\langle \cdot, \cdot \rangle$. A lattice is {\it
nondegenerate} if its form is nondegenerate.

Let $L$ be a lattice. For $m\in \Bbb{Q}$, we set $L_{m}=\{\alpha\in
L\;|\;\langle \alpha, \alpha \rangle=m\}$. The lattice $L$ is said to
be {\it even} if $L_{m}=0$ for any $m\in \Bbb{Q}$ which is not an
even integer. The lattice $L$ is said to be {\it integral} if the form
is integral-valued and to be {\it positive definite} if the form is
positive definite. Even lattices are integral. Let $L_{\Bbb{Q}}
=L\otimes_{\Bbb{Z}} \Bbb{Q}$. Then $L_{\Bbb{Q}}$ is an
$n$-dimensional vector space over $\Bbb{Q}$ in which $L$ is embedded
and the form on $L$ is extended to a symmetric $\Bbb{Q}$-bilinear
form on $L_{\Bbb{Q}}$. The lattice is nondegenerate if and only if this
form on $L_{\Bbb{Q}}$ is nondegenerate. A lattice may be equivalently
defined as the $\Bbb{Z}$-span of a basis of a finite-dimensional
rational vector space equipped with a symmetric bilinear form. The
dual of $L$ is the set $L^{\circ}=\{\alpha\in L_{\Bbb{Q}}\; |\;\langle
\alpha, L\rangle\subset \Bbb{Z}\}$. This set is a lattice if and if
$L$ is nondegenerate, and in this case, $L^{\circ}$ has as a base the
dual base of a given base. The lattice $L$ is said to be {\it
self-dual} if $L=L^{\circ}$. This is equivalent to $L$ being integral
and {\it unimodular}, which means that $|\det(\langle \alpha_{i},
\alpha_{j}\rangle)|=1$.

Recall that the Golay code $\cal{C}$ is defined on a $24$-element set
$\Omega$. Let ${\frak h}=\coprod_{k\in \Omega}\Bbb{C}\alpha_{k}$ be a
vector space with basis $\{\alpha_{k}\;|\; k\in \Omega\}$ and provided
${\frak h}$ with the symmetric bilinear form $\langle\cdot,
\cdot\rangle$ such that $\langle \alpha_{k},
\alpha_{l}\rangle=2\delta_{kl}$ for $k, l\in \Omega$.  For $S\subset
\Omega$, set $\alpha_{S}=\sum_{k\in S}\alpha_{k}$. For any fixed
element $k_{0}$ of $\Omega$, the subset
\begin{eqnarray*}
\Lambda&=&\sum_{C\in \cal{C}}\Bbb{Z}\frac{1}{2}\alpha_{C}
+\sum_{k\in \Omega}
\Bbb{Z}(\frac{1}{4}\alpha_{\Omega}-\alpha_{k})\\
&=&\sum_{C\in \cal{C}}\Bbb{Z}\frac{1}{2}\alpha_{C}+
\sum_{k, l}\Bbb{Z}(\alpha_{k}+\alpha_{l})+\Bbb{Z}
(\frac{1}{4}\alpha_{\Omega}-\alpha_{k_{0}}),
\end{eqnarray*}
of ${\frak h}$, equipped with the restriction to $\Lambda$ of the form
on ${\frak h}$, is a lattice.  This lattice is the {\it Leech
lattice}.

\begin{theo}
The Leech lattice $\Lambda$ is an even unimodular lattice such that
$\Lambda_{2}=\emptyset$. It is unique up to isometry.
\end{theo}

The Leech lattice is generated by $\Lambda_{4}$.  It is easy to see
that the elements $\pm\alpha_{k}\pm \alpha_{l}$, $k, l\in
\Omega$, $k\ne l$, are in $\Lambda_{4}$. Obviously,
$\theta: \Lambda \to \Lambda, \alpha\mapsto -\alpha$ is an isometry of
the Leech lattice such that $\theta^{2}=1$.

\subsection{The untwisted vertex operator algebra $V_{\Lambda}$}
Let ${\frak h}$ be the same vector space as in Subsection 1.2. We view ${\frak
h}$ as an abelian Lie algebra and consider the $\Bbb{Z}$-graded
untwisted affine Lie algebra $\tilde{\frak h}=\coprod_{n\in \Bbb{Z}}
{\frak h}\otimes t^{n}\oplus \Bbb{C}c\oplus \Bbb{C}d$, its
Heisenberg subalgebra $\hat{\frak h}_{\Bbb{Z}}=\coprod_{n\in \Bbb{Z},
n\ne 0}{\frak h}
\otimes t^{n}\oplus \Bbb{C}c$ and the subalgebra
$\hat{\frak h}_{\Bbb{Z}}^{-}=\coprod_{n<0}{\frak h}\otimes t^{n}$.  The
symmetric algebra $S(\hat{\frak h}_{\Bbb{Z}}^{-})$ over $\hat{\frak
h}_{\Bbb{Z}}^{-}$ is a $\Bbb{Z}$-graded $\hat{\frak h}_{\Bbb
Z}$-irreducible $\tilde{\frak h}$-module. Let $\hat{\Lambda}$ be a
central extension of $\Lambda$ by a cyclic group $\langle
\kappa\rangle$ of order $2$. We denote the projection {}from
$\hat{\Lambda}$ to $\Lambda$ by $^{-}$. Define the faithful character
$\chi:\langle \kappa\rangle \to \Bbb{C}^{\times}$ by
$\chi(\kappa)=-1$.  Denote by $\Bbb{C}_{\chi}$ the one-dimensional
space $\Bbb{C}$ viewed as a $\langle \kappa\rangle$-module on which
$\langle \kappa\rangle$ acts according to $\chi$ and denote by
$\Bbb{C}\{\Lambda\}$ the induced $\hat{\Lambda}$-module
$\Bbb{C}[\hat{\Lambda}]\otimes_{\Bbb{C} [\langle \kappa\rangle]}
\Bbb{C}_{\chi}$ (where $\Bbb{C}[\hat{\Lambda}]$ and $\Bbb{C}[\langle
\kappa\rangle]$ are the group algebras of $\hat{\Lambda}$ and $\langle
\kappa\rangle$, respectively). Set $V_{\Lambda}=S(\hat{\frak h}_{\Bbb{Z}}
^{-})\otimes \Bbb{C}\{\Lambda\}$.  We regard $S(\hat{\frak h}_{\Bbb{Z}}^{-})$
 as the trivial $\hat{\Lambda}$-module and $V_{\Lambda}$ as
the corresponding tensor product $\hat{\Lambda}$-module. View
$\Bbb{C}\{\Lambda\}$ as a trivial $\hat{\frak h}_{\Bbb{Z}}$-module and for
$\alpha\in {\frak h}$, define $\alpha(0): \Bbb{C}\{\Lambda\}\to
\Bbb{C}\{\Lambda\}$ by $\alpha(0)(a\otimes 1)=\langle \alpha,
\bar{a}\rangle(a\otimes 1)$ for any $a\in
\hat{\Lambda}$. Also define
$x^{\alpha}\in (\mbox{\rm End}\ \Bbb{C}\{\Lambda\})[x, x^{-1}]$ for
$\alpha\in
\Lambda$ by $x^{\alpha}(a\otimes 1)=x^{\langle \alpha,
\bar{a}\rangle}(a\otimes 1)$ for any $a\in \hat{\Lambda}$. Give
$\Bbb{C}\{\Lambda\}$ a $\Bbb{Z}$-gradation ({\it weight}) by $\wt
a\otimes 1=
\frac{1}{2}\langle \bar{a}, \bar{a}\rangle$ for all $a\in
\hat{\Lambda}$. Then
$V_{\Lambda}$ has a $\Bbb{Z}$-gradation ({\it weight}) obtained
{}from the tensor product gradation. Let $d\in \tilde{\frak h}$ act
as the weight operators on $S(\hat{\frak h}_{\Bbb{Z}}^{-})$ and on
$\Bbb{C}\{\Lambda\}$ and give $V_{\Lambda}$ the tensor product
$\tilde{\frak h}$-module structure. We denote the action of
$\alpha\otimes t^{n}$ by $\alpha(n)$ for $\alpha\in {\frak h}$ and
denote the element $1\otimes (a\otimes 1)\in V_{\Lambda}$ by
$\iota(a)$ for $a\in \hat{\Lambda}$.

For $\alpha\in {\frak h}$, let $\alpha(x)=
\sum_{n\in \Bbb{Z}}\alpha(n)x^{-n-1}$.  For any $a\in \hat{\Lambda}$, we define
\begin{eqnarray*}
Y_{V_{\Lambda}}(\iota(a), x)&=&\nord e^{\int \bar{a}(x)}\nord\\
&=&\nord \exp\left(-\sum_{n<0}\frac{\bar{a}(n)}{n}x^{-n}\right)
\exp\left(-\sum_{n>0}\frac{\bar{a}(n)}{n}x^{-n}\right)ax^{\bar{a}}\nord
\end{eqnarray*}
(the {\it (untwisted) vertex operator associated to $\iota(a)$}) of
$(\mbox{\rm End}\ V_{\Lambda})[[x, x^{-1}]]$, where the normal
ordering is defined by
\begin{eqnarray*}
\nord \alpha_{1}(m)\alpha_{2}(n)\nord&=&\nord \alpha_{2}(n)
\alpha_{1}(m)\nord
=\left\{\begin{array}{ll}
\alpha_{1}(m)\alpha_{2}(n)&m\le n,\\
\alpha_{2}(n)\alpha_{1}(m)&m\ge n,\end{array}\right.\\
\nord \alpha(m)a\nord&=&\nord a\alpha(m)\nord=a\alpha(m),\\
\nord x^{\alpha}a\nord&=&\nord ax^{\alpha}\nord=ax^{\alpha}
\end{eqnarray*}
for $\alpha_{1}, \alpha_{2}, \alpha\in {\frak h}$, $m, n\in \Bbb{Z}$,
$\alpha\in \Lambda$ and $a\in \hat{\Lambda}$. For
$v=\alpha_{1}(-n_{1})\cdots \alpha_{k}(n_{k})\cdot \iota(a)$, we
define the {\it (untwisted) vertex operator} associated to $v$ to be
\begin{eqnarray*}
Y_{V_{\Lambda}}(v, x)&=&\nord \left(\frac{1}{(n_{1}-1)!}
\frac{d^{n_{1}-1}
\alpha_{1}(x)}{dx^{n_{1}-1}}\right)\cdots \\
&&\hspace{4em}\cdot \left(\frac{1}{(n_{k}-1)!}\frac{d^{n_{k}-1}
\alpha_{k}(x)}{dx^{n_{k}-1}}\right)Y_{V_{\Lambda}}(\iota(a), x)\nord.
\end{eqnarray*}
Extending $Y_{V_{\Lambda}}$ by linearity, we obtain a linear map
$V_{\Lambda}\to (\mbox{\rm End}\ V_{\Lambda})[[x, x^{-1}]]$, $v\to
Y_{V_{\Lambda}}(v, x)$. Let $\{h_{1}, \dots, h_{24}\}$ be an
orthonormal basis of ${\frak h}$. Consider
$\omega=\frac{1}{2}\sum_{i=1}^{24}h_{i}(-1)^{2}$.  The following
theorem is a special case of a theorem due to Borcherds \cite{B1};
see \cite{FLM2}:

\begin{theo}
The quadruple $(V_{\Lambda}, Y_{V_{\Lambda}}, \iota(1), \omega)$ is a
vertex operator algebra with central charge (or rank) equal to 24.
\end{theo}

When there is no confusion, we shall use $Y$ to denote the vertex
operator map $Y_{V_{\Lambda}}$.

The following theorem is a special case of a theorem due to Dong
\cite{D1}:

\begin{theo}
Any irreducible $V_{\Lambda}$-module is isomorphic to $V_{\Lambda}$ as
a $V_{\Lambda}$-module and any $V_{\Lambda}$-module is a finite sum of
copies of $V_{\Lambda}$ as a $V_{\Lambda}$-module.
\end{theo}

On $\Bbb{C}\{\Lambda\}$ there is a unique positive definite hermitian
form $(\cdot, \cdot)_{\Bbb{C}\{\Lambda\}}$ (see \cite{FLM2}) such
that $$(a\otimes 1, b\otimes 1)_{\Bbb{C}\{\Lambda\}}
=\left\{\begin{array}{ll} 0&
\bar{a}\ne \bar{b},\\ 1&a=b.\end{array}\right.$$
It is easy to see that on $V_{\Lambda}$ there is a unique bilinear
form $\langle \cdot ,
\cdot \rangle_{V_{\Lambda}}$ such that
\begin{eqnarray*}
\langle \iota(a), \iota(b)\rangle_{V_{\Lambda}}
&=&(a\otimes 1, b\otimes 1)_{\Bbb{C}\{\Lambda\}},\\
\langle d\cdot u, v\rangle_{V_{\Lambda}}&
=&\langle  u, d\cdot v\rangle_{V_{\Lambda}},\\
\langle \alpha(n)\cdot u, v\rangle_{V_{\Lambda}}&
=&\langle  u, \alpha(n)\cdot v\rangle_{V_{\Lambda}}.
\end{eqnarray*}

Recall the isometry $\theta$ of the Leech lattice. For
$v=\alpha_{1}(-n_{1})\cdots \alpha_{k}(-n_{k})\cdot \iota(a)\in
V_{\Lambda}$, we define $\theta(v)=(-1)^{k}\alpha_{1}(-n_{1})\cdots
\alpha_{k}(-n_{k})\cdot \iota(a^{-1})$.  Using linearity, we obtain a
linear map $\theta: V_{\Lambda}\to V_{\Lambda}$.  It is clear that
$\theta^{2}=1$ and $\theta$ is an automorphism of the vertex operator
algebra $V_{\Lambda}$. Thus $V_{\Lambda}=V_{\Lambda}^{+}+
V_{\Lambda}^{-}$ where $V_{\Lambda}^{\pm}$ are the eigenspace of
$\theta$ with eigenvalue $\pm 1$. The subspace $V_{\Lambda}^{+}$ is a
vertex operator algebra and both $V_{\Lambda}^{\pm}$ are irreducible
$V_{\Lambda}^{+}$-modules.

\subsection{The twisted module $V_{\Lambda}^{T}$ for
$V_{\Lambda}$} Consider the $\Bbb{Z}+\frac{1}{2}$-graded twisted
affine Lie algebra $\tilde{\frak h}[-1]
=\coprod_{n\in \Bbb{Z}+\frac{1}{2}} {\frak h}\otimes
t^{n}\oplus \Bbb{C}c\oplus \Bbb{C}d$,
its Heisenberg subalgebra $\hat{\frak h}_{\Bbb{Z}+\frac{1}{2}}
=\coprod_{n\in \Bbb{Z}+\frac{1}{2}}{\frak h}
\otimes t^{n}\oplus \Bbb{C}c$ and the subalgebra
$\hat{\frak h}_{\Bbb{Z}+\frac{1}{2}}^{-}=\coprod_{n<0}{\frak
h}\otimes t^{n}$.  The symmetric algebra
$S(\hat{\frak h}_{\Bbb{Z}+\frac{1}{2}}^{-})$ over
$\hat{\frak h}_{\Bbb{Z}+\frac{1}{2}}^{-}$
is a $\Bbb{Z}+\frac{1}{2}$-graded
$\hat{\frak h}_{\Bbb{Z}+\frac{1}{2}}$-irreducible $\tilde{\frak h}[-1]$-module.

Let $K=\{a^{2}\kappa^{\langle \bar{a}, \bar{a}\rangle/2}\;|\;a\in
\hat{\Lambda}\}$.  Then $K$ is a central subgroup of $\hat{\Lambda}$.
The following result is proved in \cite{FLM2}:

\begin{theo}
The quotient group $\hat{\Lambda}/K$ has a unique (up to equivalence)
irreducible module $T$ such that $\kappa K\to -1$ on $T$. Moreover,
the corresponding representation of $\hat{\Lambda}/K$ is the unique
faithful irreducible representation and $\dim T=2^{12}$. To construct
$T$, let $\Phi$ be any subgroup of $\Lambda$ such that
$2\Lambda\subset \Phi \subset
\Lambda$, $|\Phi/2\Lambda|=2^{12}$ and $\frac{1}{2}\langle \alpha,
\alpha\rangle\in 2\Bbb{Z}$ for any $\alpha\in \Phi$. Then the
preimage
$\hat{\Phi}$ of $\Phi$ under the homomorphism $^{-}: \hat{\Lambda}\to
\Lambda$ is a maximal subgroup of $\hat{\Lambda}$ and $\hat{\Phi}/K$
is an elementary abelian $2$-group. Let $\Psi: \hat{\Phi}/K\to
\Bbb{C}^{\times}$ be any homomorphism such that $\Psi(\kappa K)=-1$ and
denote by $\Bbb{C}_{\Psi}$ the one-dimensional $\hat{\Phi}$-module
with the corresponding character. Then viewed as a
$\hat{\Lambda}$-module
\begin{eqnarray*}
T&=&\Bbb{C}[\hat{\Lambda}]\otimes_{\Bbb{C}[\hat{\Phi}]}
\Bbb{C}_{\Psi}\nno\\ &\simeq& \Bbb{C}[\Lambda/\Phi]\;\;\;\;\;\mbox{\it
(linearly)}.
\end{eqnarray*}
\end{theo}

For any $a\in \hat{\Lambda}$, the element $a\otimes 1\in T$ is denoted
by $t(a)$.

Set $V^{T}_{\Lambda}=S(\hat{\frak h}_{\Bbb{Z}+\frac{1}{2}}^{-})\otimes T$.
We view $T$ as a
$\hat{\Lambda}$-module and regard $S(\hat{\frak h}_{\Bbb{Z}+\frac{1}{2}}^{-})$
as the trivial $\hat{\Lambda}$-module and
$V_{\Lambda}$ as the corresponding tensor product
$\hat{\Lambda}$-module. Give $T$ a $\Bbb{Z}+\frac{1}{2}$-gradation
({\it weight}) by $\wt t=
\frac{24}{16}=\frac{3}{2}$ for all $t\in T$. Then
$V_{\Lambda}^{T}$ has a $\Bbb{Z}+\frac{1}{2}$-gradation ({\it
weight}) obtained {}from the tensor product gradation.  Let $d\in
\tilde{\frak h}$ act as the weight operators on $S(\hat{\frak
h}_{\Bbb{Z}+\frac{1}{2}}^{-})$ and on $T$.  View $T$ as a trivial
$\hat{\frak h}$-module and give $V_{\Lambda}$ the tensor product
$\tilde{\frak h}[-1]$-module structure.  We denote the action of
$\alpha\otimes t^{n}$ for $\alpha\in {\frak h}$ and $n\in
\Bbb{Z}+\frac{1}{2}$ by $\alpha(n)$.

For any $\alpha\in {\frak h}$, let $\alpha(x) =\sum_{n\in
\Bbb{Z}+\frac{1}{2}}\alpha(n)x^{-n-1}$. (Note that though we use the same
notation as in Subsection 1.3, $\alpha(x)$ in Subsection 1.3 acts on a
different space.) For any $a\in \hat{\Lambda}$, we define the {\it
twisted vertex operator}
\begin{eqnarray*}
Y_{V_{\Lambda}^{T}}(\iota(a), x)&=&2^{-\langle \bar{a},
\bar{a}\rangle}
\nord e^{\int \bar{a}(x)}\nord ax^{-\langle \bar{a},
\bar{a}\rangle/2}\\
&=&2^{-\langle \bar{a}, \bar{a}\rangle}\exp\left(\sum_{n< 0}
\frac{\bar{a}(n+\frac{1}{2})}{n+\frac{1}{2}}
x^{-(n+\frac{1}{2})}\right)\cdot\\
&&\hspace{3em}\cdot\exp\left(\sum_{n\ge 0}
\frac{\bar{a}(n+\frac{1}{2})}{n+\frac{1}{2}}x^{-(n+\frac{1}{2})}\right)
ax^{-\langle \bar{a}, \bar{a}\rangle/2}
\end{eqnarray*}
of $(\mbox{\rm End}\ V_{\Lambda}^{T})[[x^{\frac{1}{2}}, x^{-\frac{1}{2}}]]$,
where the normal ordering is defined by
\begin{eqnarray*}
\nord \alpha_{1}(m)\alpha_{2}(n)\nord&=&\nord \alpha_{2}(n)
\alpha_{1}(m)\nord
=\left\{\begin{array}{ll}
\alpha_{1}(m)\alpha_{2}(n)&m\le n,\\
\alpha_{2}(n)\alpha_{1}(m)&m\ge n,\end{array}\right.
\end{eqnarray*}
for $\alpha_{1}, \alpha_{2}\in {\frak h}$, $m, n\in \Bbb{Z}$.
For $v=\alpha_{1}(-n_{1})\cdots \alpha_{k}(-n_{k})\cdot \iota(a)
\in V_{\Lambda}$,
we define
\begin{eqnarray*}
Y_{0}(v, x)&=&\nord \left(\frac{1}{(n_{1}-1)!}\frac{d^{n_{1}-1}
\alpha_{1}(x)}{dx^{n_{1}-1}}\right)\cdots \\
&&\hspace{3em}\cdot \left(\frac{1}{(n_{k}-1)!}
\frac{d^{n_{k}-1}
\alpha_{k}(x)}{dx^{n_{k}-1}}\right)Y_{V_{\Lambda}^{T}}(\iota(a),
x)\nord.
\end{eqnarray*}
Let $c_{mn}$ be the complex numbers determined by the formula
$$\sum_{n, m\ge 0}c_{mn}x^{m}y^{n} =-\log
\left(\frac{(1+x)^{1/2}+(1+y)^{1/2}}{2}\right).$$ We define the {\it
twisted vertex operator associated to $v$} to be
$$Y_{V_{\Lambda}^{T}}(v, x)=Y_{0}\left(\exp\left(\sum_{m, n\ge 0}
\sum_{i=1}^{24}c_{mn}h_{i}(m)h_{i}(n)x^{-m-n}\right)v, x\right)$$
where as in
Subsection 1.3 $\{h_{1}, \dots, h_{24}\}$ is an orthogonal basis for
${\frak h}$. The following theorem is a special case of a theorem due
Frenkel, Lepowsky and Meurman \cite{FLM2}:

\begin{theo}
The pair $(V_{\Lambda}^{T}, Y_{V_{\Lambda}^{T}})$ is a
$\theta$-twisted $V_{\Lambda}$-module.
\end{theo}

When there is no confusion,  we shall use $Y$ to denote the
vertex operator map $Y_{V_{\Lambda}^{T}}$.

The following theorem is a special case of a theorem due to Dong
\cite{D2}:

\begin{theo}
Any irreducible $\theta$-twisted $V_{\Lambda}$-module is isomorphic to
$V_{\Lambda}^{T}$ and any $\theta$-twisted $V_{\Lambda}$-module is a
finite sum of copies of $V_{\Lambda}^{T}$.
\end{theo}

On $T$ there is a unique positive definite hermitian form $(\cdot,
\cdot)_{T}$ (see \cite{FLM2}) such that $$(t(a),
t(b))_{T}=\left\{\begin{array}{ll} 0& a\hat{\Phi}\ne b\hat{\Phi},\\
1&a=b.\end{array}\right.$$ Thus on $V_{\Lambda}$ there is a unique
bilinear form $\langle \cdot ,
\cdot \rangle_{V_{\Lambda}^{T}}$ such that
\begin{eqnarray*}
\langle 1\otimes t(a), 1\otimes t(b)\rangle_{V_{\Lambda}^{T}}
&=&(t(a), t(b))_{T},\\
\langle d\cdot u, v\rangle_{V_{\Lambda}^{T}}&
=&\langle  u, d\cdot v\rangle_{V_{\Lambda}^{T}},\\
\langle \alpha(n)\cdot u, v\rangle_{V_{\Lambda}^{T}}&
=&\langle  u, \alpha(-n)\cdot v\rangle_{V_{\Lambda}^{T}}.
\end{eqnarray*}

 For $w=\alpha_{1}(-n_{1})\cdots \alpha_{k}(-n_{k})\otimes t\in
V_{\Lambda}^{T}$, we define
$$\theta(w)=(-1)^{k+1}\alpha_{1}(-n_{1})\cdots
\alpha_{k}(-n_{k})\otimes t.$$
Using linearity, we obtain a linear map $\theta: V_{\Lambda}^{T}\to
V_{\Lambda}^{T}$.  It is clear that $\theta^{2}=1$ and $\theta$ is an
automorphism of the twisted $V_{\Lambda}$-module $V_{\Lambda}^{T}$.
Thus $V_{\Lambda}^{T} =(V_{\Lambda}^{T})^{+}+ (V_{\Lambda}^{T})^{-}$
where $(V_{\Lambda}^{T})^{\pm}$ are the eigenspace of $\theta$ with
eigenvalue $\pm 1$. Both $(V_{\Lambda}^{T})^{+}$ and
$(V_{\Lambda}^{T})^{-}$ are irreducible $V_{\Lambda}^{+}$-modules.

\subsection{The moonshine module $V^{\natural}$}
The {\it moonshine module} is defined to be
$V^{\natural}=V_{\Lambda}^{+}\oplus (V_{\Lambda}^{T})^{+}$. It is not
difficult to show that the generating function of the dimensions of
the homogeneous subspaces of $V^{\natural}$ is equal to
$qJ(q)=q(j(q)-744)$.  The following result is established in
\cite{FLM2}:

\begin{theo}\label{monster}
The $\Bbb{Z}$-graded space $V^{\natural}$ has a natural vertex
operator algebra structure and its automorphism group is the Monster.
\end{theo}

The proof of Theorem \ref{monster} in \cite{FLM2}
uses triality and
some results in group theory.
Though the proof of the theorem above is involved, there is a direct
and natural way to define the vertex operator map $Y_{V^{\natural}}$
as was carried out in \cite{FHL} for the sum of an arbitrary vertex
operator algebra and an arbitrary $\Bbb{Z}$-graded module for the
vertex operator algebra. We recall this construction in the case of
the moonshine module here: For $u, v\in V_{\Lambda}^{+}$,
$Y_{V^{\natural}}(u, x)v=Y_{V_{\Lambda}}(u, x)v$; for $u\in
V_{\Lambda}^{+}$, $v\in (V_{\Lambda}^{T})^{+}$, $Y_{V^{\natural}}(u,
x)v =Y_{V_{\Lambda}^{T}}(u, x)v$; for $u\in (V_{\Lambda}^{T})^{+}$,
$v\in V_{\Lambda}^{+}$, $Y_{V^{\natural}}(u, x)v
=e^{xL(-1)}Y_{V_{\Lambda}^{T}}(v, -x)u$; for $u, v\in
(V_{\Lambda}^{T})^{+}$, $Y_{V^{\natural}}(u, x)v$ is defined by
$$\langle w, Y_{V^{\natural}}(u, x)v\rangle_{V_{\Lambda}} =\langle
Y_{V_{\Lambda}^{T}}(w, -x^{-1})e^{xL(1)}(-x^{2})^{-L(0)}u,
e^{x^{-1}L(1)}v\rangle_{V_{\Lambda}^{T}}$$
for all $w\in V_{\Lambda}$. The vacuum of
$V^{\natural}$ is $\iota(1)$ and the Virasoro element is $\omega$.

We now discuss the decompositions of $V_{\Lambda}^{+}$ and its modules
into direct sums of modules for a tensor product of the Virasoro
vertex operator algebra of central charge $\frac{1}{2}$.  Let
$L(\frac{1}{2}, 0)$ be the rational Virasoro vertex operator algebra
of central charge $\frac{1}{2}$ and $L(\frac{1}{2}, i)$, $i=0,
\frac{1}{2}, \frac{1}{16}$, the irreducible $L(\frac{1}{2},
0)$-modules (see \cite{FZ} and \cite{DMZ}).  The following result is
proved by Dong, Mason and Zhu \cite{DMZ}:

\begin{theo}\label{dmz}
There exist $\omega_{i}\in V_{\Lambda}^{+}$, $i=1, \dots, 48$, such
that $\omega=\omega_{1}+\cdots +\omega_{48}$ and the vertex operator
subalgebra $L$ of $V_{\Lambda}^{+}$ generated by $\omega_{i}$, $i=1,
\dots, 48$, is isomorphic to $L(\frac{1}{2}, 0)^{\otimes 48}$. In
particular, for $W=V_{\Lambda}^{\pm}, (V_{\Lambda}^{T})^{\pm}$, $(W,
Y_{W}\mbox{\huge $\vert$}_{L\otimes W})$ are  $L(\frac{1}{2}, 0)^{\otimes
48}$-modules.
\end{theo}

A vector in an $L(\frac{1}{2}, 0)^{48}$-module $W$ is a {\it lowest
weight vector} if it is a lowest weight vector when $W$ is regarded as
an $L(\frac{1}{2}, 0)$-module for any one of the $48$ vertex operator
subalgebras $L(\frac{1}{2}, 0)$. The {\it lowest weight} of a
lowest vector $v$ is
defined in the obvious way. It is an array of $48$ complex numbers.
For any homogeneous subspace $W_{(n)}$ of an $L(\frac{1}{2},
0)^{48}$-module $W$, we denote the subspace spanned by all lowest
weight vectors in $W_{(n)}$ by $W_{(n)}^{l}$. Let $a, b, c$ be
nonnegative integers such that $a+b+c=48$. A lowest weight vector
$v\in W$ with lowest weight $(h_{1},
\dots, h_{48})$ is called a vector of type $(a, b, c)$ if $\#
\{h_{i}\:|\: h_{i}=0\}=a$, $\# \{h_{i}\:|\:h_{i}=\frac{1}{2}\}=b$
and $\# \{h_{i}\:|\: h_{i}=\frac{1}{16}\}=c$. We write
$W_{(n)}^{l}=\coprod_{a, b, c}m_{a, b, c}(a, b, c)$ which means that
there are $m_{a, b, c}$ linearly independent vectors of type $(a, b,
c)$ in $W_{(n)}$. In \cite{DMZ}, the following information on the
lowest weight vectors in $V_{\Lambda}$ and in $V_{\Lambda}^{T}$ is
obtained:

\begin{theo}\label{decomp}
We have the following decompositions:
\begin{eqnarray*}
(V_{\Lambda})_{(1)}^{l}&=&24(46, 2, 0),\\
(V_{\Lambda}^{T})_{(3/2)}^{l}&=&2^{12}(24, 0, 24),\\
(V_{\Lambda}^{T})_{(2)}^{l}&=&24\cdot 2^{12}(23, 1, 24).
\end{eqnarray*}
\end{theo}

The following information on the decompositions of $V_{\Lambda}^{\pm},
(V_{\Lambda}^{T})^{\pm}$ into direct sums of irreducible
$L(\frac{1}{2}, 0)^{\otimes 48}$-modules can be found in \cite{D3}:

\begin{theo}\label{d3}
Let $W=V_{\Lambda}^{\pm}, (V_{\Lambda}^{T})^{\pm}$.  As an
$L(\frac{1}{2}, 0)^{\otimes 48}$-module, $$W=\coprod_{h_{i}=0,
\frac{1}{2}, \frac{1}{16}}c_{h_{1}\cdots h_{48}} L(\frac{1}{2},
h_{1})\otimes \cdots \otimes L(\frac{1}{2}, h_{48}).$$ If
$W=V_{\Lambda}^{\pm}$ and $c_{h_{1}\cdots h_{48}}\ne 0$ for $h_{i}\in
\{0, \frac{1}{2}, \frac{1}{16}\}$, $i=1, \dots, 48$, then $$(h_{2j-1},
h_{2j})\in \{(0, 0), (0, \frac{1}{2}), (\frac{1}{2}, 0), (\frac{1}{2},
\frac{1}{2}), (\frac{1}{16}, \frac{1}{16})\},$$ $1\le j\le 24$.  If
$W=(V_{\Lambda}^{T})^{\pm}$ and $c_{h_{1}\cdots h_{48}}\ne 0$ for
$h_{i}\in \{0, \frac{1}{2}, \frac{1}{16}\}$, $i=1, \dots, 48$, then
$$(h_{2j-1}, h_{2j})\in \{(0, \frac{1}{16}), (\frac{1}{16}, 0),
(\frac{1}{2}, \frac{1}{16}), (\frac{1}{16}, \frac{1}{2})\},$$ $1\le
j\le 24$.  Let $x_{1}^{i}, \dots, x_{48}^{i}$, $i=1, \dots, 24$, be
given by $(x_{2j-1}^{i}, x^{i}_{2j})=(0, 0)$, $j=1, \dots, 24$, $j\ne
i$, and $(x_{2j-1}^{j}, x_{2j}^{j})=(\frac{1}{2}, \frac{1}{2})$, $j=1,
\dots, 24$.  When $W=V_{\Lambda}^{-}$, the multiplicities
$c_{x_{1}^{i}\cdots x_{48}^{i}}=1$, $i=1, \dots, 24$.
\end{theo}

The following results are due to Dong \cite{D3}:

\begin{theo}\label{do1}
The vertex operator algebra $V_{\Lambda}^{+}$ has only the four
irreducible modules $V_{\Lambda}^{\pm}, (V_{\Lambda}^{T})^{\pm}$ (up
to isomorphism) and any $V_{\Lambda}^{+}$-module is a finite sum of
irreducible modules.
\end{theo}

\begin{theo}
The moonshine module vertex operator algebra $V^{\natural}$ has only
one irreducible module, $V^{\natural}$ itself, (up to isomorphisms) and any
$V^{\natural}$-module is a finite sum of irreducible modules.
\end{theo}

\section{Tensor products of modules for a vertex operator algebra}

In this section we summarize the basic concepts and constructions in
the theory of tensor products of modules for a vertex operator algebra
and those results (mainly the associativity) which we need in this
paper. Details can be found in
\cite{HL2}--\cite{HL3}, \cite{HL5}, \cite{H2}.
This theory is initiated in \cite{HL1}.
For the complete picture of the tensor
product theory, see \cite{HL4}.

\subsection{The definition, some properties and two constructions
of $P(z)$-tensor products} Let $(V, Y, \bold{1}, \omega)$ be a vertex
operator algebra and $(W, Y)$ a $V$-module.  For any $v\in V$ and
$n\in \Bbb{Z}$, there is a well-defined natural action of $v_{n}$ on
$\overline{W}$.  Moreover, for fixed $v\in V$, any infinite linear
combination of the $v_{n}$ of the form $\sum_{n<N}a_{n}v_{n}$
($a_{n}\in \Bbb{C}$) acts on $\overline{W}$ in a well-defined way.

Fix $z\in \Bbb{C}^{\times}$ and let $(W_{1}, Y_{1})$, $(W_{2},
Y_{2})$ and $(W_{3}, Y_{3})$ be $V$-modules.  A {\it
$P(z)$-intertwining map of type ${W_{3}}\choose {W_{1}W_{2}}$} is a
linear map $F: W_{1}\otimes W_{2} \to
\overline{W}_{3}$ satisfying the condition
\begin{eqnarray*}
\lefteqn{x_{0}^{-1}\delta\left(\frac{ x_{1}-z}{x_{0}}\right)
Y_{3}(v, x_{1})F(w_{(1)}\otimes w_{(2)})=}\nonumber\\
&&=z^{-1}\delta\left(\frac{x_{1}-x_{0}}{z}\right)
F(Y_{1}(v, x_{0})w_{(1)}\otimes w_{(2)})\nonumber\\
&&\hspace{2em}+x_{0}^{-1}\delta\left(\frac{z-x_{1}}{-x_{0}}\right)
F(w_{(1)}\otimes Y_{2}(v, x_{1})w_{(2)})
\end{eqnarray*}
for $v\in V$, $w_{(1)}\in W_{1}$, $w_{(2)}\in W_{2}$.  The vector
space of $P(z)$-intertwining maps of type ${W_{3}}\choose
{W_{1}W_{2}}$ is denoted by $\cal{M}[P(z)]^{W_{3}}_{W_{1}W_{2}}$.  A
{\it $P(z)$-product of $W_{1}$ and $W_{2}$} is a $V$-module $(W_{3},
Y_{3})$ equipped with a $P(z)$-intertwining map $F$ of type
${W_{3}}\choose {W_{1}W_{2}}$ and is denoted by $(W_{3}, Y_{3}; F)$
(or simply by $(W_{3}, F)$).  Let $(W_{4}, Y_{4}; G)$ be another
$P(z)$-product of $W_{1}$ and $W_{2}$.  A {\it morphism} {}from
$(W_{3}, Y_{3}; F)$ to $(W_{4}, Y_{4}; G)$ is a module map $\eta$
{}from $W_{3}$ to $W_{4}$ such that $G=\overline{\eta}\circ F$, where
$\overline{\eta}$ is the map {}from $\overline{W}_{3}$ to
$\overline{W}_{4}$ uniquely extending $\eta$.

A {\it $P(z)$-tensor product of $W_{1}$ and $W_{2}$} is a
$P(z)$-product $$(W_{1}\boxtimes_{P(z)} W_{2}, Y_{P(z)};
\boxtimes_{P(z)})$$
such that for any $P(z)$-product
$(W_{3}, Y_{3}; F)$, there is a unique morphism {}from
$$(W_{1}\boxtimes_{P(z)} W_{2}, Y_{P(z)};
\boxtimes_{P(z)})$$ to $(W_{3}, Y_{3}; F)$.
The $V$-module $(W_{1}\boxtimes_{P(z)} W_{2},  Y_{P(z)})$ is
called a {\it $P(z)$-tensor product module} of $W_{1}$ and $W_{2}$.
A $P(z)$-tensor product is unique up to isomorphism.

We have the following properties:

\begin{propo}\label{op-map}
Let $\log z=\log |z|+i \arg z$ such that $0\le \arg z < 2 \pi$ and
$l_{p}(z)=\log z+2\pi p i$, $p\in \Bbb{Z}$. For any value $p\in
\Bbb{Z}$, we have an isomorphism {}from the vector space ${\cal
V}_{W_{1}W_{2}}^{W_{3}}$ of intertwining operators of type
${W_{3}}\choose {W_{1}W_{2}}$ to the vector space ${\cal
M}[P(z)]_{W_{1}W_{2}}^{W_{3}}$. This isomorphism takes an intertwining
operator of the type ${W_{3}}\choose {W_{1}W_{2}}$ to the
$P(z)$-intertwining map of the same type obtained {}from the
intertwining operator by substituting the complex powers of
$e^{l_{p}(z)}$ for the complex powers of the formal variable.
\end{propo}

\begin{propo}
Suppose that $W_{1}\boxtimes_{P(z)}W_{2}$ exists. We have a natural
isomorphism
\begin{eqnarray}
\mbox{\normalshape Hom}_{V}(W_{1}\boxtimes_{P(z)}W_{2}, W_{3})&
\stackrel{\sim}{\to}&
\cal{M}[P(z)]^{W_{3}}_{W_{1}W_{2}}\nno\\
\eta&\mapsto& \overline{\eta}\circ \boxtimes_{P(z)}.
\end{eqnarray}
\end{propo}

\begin{propo}
Let $U_{1}, \dots, U_{k}$, $W_{1}, \dots, W_{l}$ be $V$-modules and
suppose that each $U_{i}\boxtimes_{P(z)}W_{j}$ exists. Then
$(\coprod_{i}U_{i})\boxtimes_{P(z)}(\coprod_{j}W_{j})$ exists and
there is a natural isomorphism
\begin{equation}
\biggl(\coprod_{i}U_{i}\biggr)\boxtimes_{P(z)}\biggl(\coprod_{j}
W_{j}\biggr)
\stackrel{\sim}
{\rightarrow} \coprod_{i,j}U_{i}\boxtimes_{P(z)}W_{j}.
\end{equation}
\end{propo}

We consider the following  special but important class of
vertex operator algebras:
A  vertex operator algebra $V$ is {\it rational} if it
satisfies the following conditions:
\begin{enumerate}
\item There are only finitely many irreducible $V$-modules
(up to equivalence).
\item Every $V$-module is completely reducible
(and is in particular a {\it finite} direct sum of irreducible
modules).
\item All the fusion rules (the dimensions of spaces of
intertwining operators) for $V$ are finite (for triples of irreducible
modules and hence arbitrary modules).
\end{enumerate}

\begin{propo}
Let $V$ be rational and let $W_{1}$, $W_{2}$ be $V$-modules. Then
$$(W_{1}\boxtimes_{P(z)}W_{2}, Y_{P(z)}; \boxtimes_{P(z)})$$ exists
and the $P(z)$-tensor product module $W_{1}\boxtimes_{P(z)}W_{2}$ of
$W_{1}$ and $W_{2}$ is isomorphic to the $V$-module $\coprod_{i=1}^{k}
(\cal{V}^{M_{i}}_{W_{1}W_{2}})^{*}\otimes M_{i}$ where $\{ M_{1},
\dots, M_{k}\}$ is a set of representatives of the equivalence
classes of irreducible $V$-modules.
\end{propo}

We now describe the constructions of a $P(z)$-tensor product of two
modules.  For two $V$-modules $(W_{1}, Y_{1})$ and $(W_{2}, Y_{2})$,
we define an action of $$V \otimes \iota_{+}\Bbb{C}[t,t^{- 1},
(z^{-1}-t)^{-1}]$$ on $(W_1 \otimes W_2)^*$ (where as in \cite{FLM2}
and
\cite{HL2}, $\iota_{+}$ denotes the operation of expansion of a
rational function of $t$ in the direction of positive powers of $t$),
that is, a linear map
\begin{equation}
\tau_{P(z)}: V\otimes \iota_{+}\Bbb{C}[t, t^{-1},
(z^{-1}-t)^{-1}]\to
\mbox{\rm End}\;(W_{1}\otimes W_{2})^{*},
\end{equation}
by
\begin{eqnarray}
\lefteqn{\left(\tau_{P(z)}
\left(x_{0}^{-1}\delta\left(\frac{x^{-1}_{1}-z}{x_{0}}\right)
Y_{t}(v, x_{1})\right)\lambda\right)(w_{(1)}\otimes
w_{(2)})}\nonumber\\
&&=z^{-1}\delta\left(\frac{x^{-1}_{1}-x_{0}}{z}\right)
\lambda(Y_{1}(e^{x_{1}L(1)}(-x_{1}^{-2})^{L(0)}v, x_{0})w_{(1)}
\otimes w_{(2)})
\nonumber\\
&&\hspace{2em}+x^{-1}_{0}\delta\left(\frac{z-x^{-1}_{1}}{-x_{0}}
\right)
\lambda(w_{(1)}\otimes Y_{2}^{*}(v, x_{1})w_{(2)})
\end{eqnarray}
for $v\in V$, $\lambda\in (W_{1}\otimes W_{2})^{*}$, $w_{(1)}\in
W_{1}$, $w_{(2)}\in W_{2}$, where
\begin{equation}
Y_{t}(v, x)=v\otimes x^{-1}\delta\left(\frac{t}{x}\right).
\end{equation}
There is an obvious action of $$V \otimes \iota_{+}\Bbb{C}[t,t^{- 1},
(z^{-1}-t)^{-1}]$$ on any $V$-module. We have:

\begin{propo}\label{intwact}
Under the natural isomorphism
\begin{equation}
\mbox{\normalshape Hom}(W'_{3}, (W_{1}\otimes
W_{2})^{*})\stackrel{\sim}{\to}\mbox{\normalshape Hom}(W_{1}\otimes W_{2},
\overline{W}_{3}),
\end{equation}
the maps in $\mbox{\normalshape Hom}(W'_{3}, (W_{1}\otimes
W_{2})^{*})$ intertwining the two actions of
$$V \otimes \iota_{+}\Bbb{C}[t,t^{-
1},(z^{-1}-t)^{-1}]$$ on $W'_{3}$ and
$(W_{1}\otimes W_{2})^{*}$ correspond exactly to the
$P(z)$-intertwining maps of type ${W_{3}}\choose {W_{1}W_{2}}$.
\end{propo}

Write
\begin{equation}
Y'_{P(z)}(v, x)=\tau_{P(z)}(Y_{t}(v, x))
\end{equation}
and
\begin{equation}
Y'_{P(z)}(\omega, x)=\sum_{n\in \Bbb{Z}}L'_{P(z)}(n)x^{-n-2}.
\end{equation}
We call the eigenspaces of the operator $L'_{P(z)}(0)$ the {\it weight
subspaces} or {\it homogeneous subspaces} of $(W_{1}\otimes
W_{2})^{*}$, and we have the corresponding notions of {\it weight
vector} (or {\it homogeneous vector}) and {\it weight}.

Let $W$ be a subspace of $(W_{1}\otimes W_{2})^{*}$.  We say that $W$
is {\it compatible for $\tau_{P(z)}$} if every element of $W$
satisfies the following nontrivial and subtle condition (called the
{\it compatibility condition}) on $\lambda
\in (W_{1}\otimes W_{2})^{*}$: The formal Laurent series $Y'_{P(z)}(v,
x_{0})\lambda$ involves only finitely many negative powers of $x_{0}$
and
\begin{eqnarray}
\lefteqn{\tau_{P(z)}\left(x_{0}^{-1}
\delta\left(\frac{x^{-1}_{1}-z}{x_{0}}
\right)
Y_{t}(v, x_{1})\right)\lambda=}\nno\\
&&=x_{0}^{-1}\delta\left(\frac{x^{-1}_{1}-z}{x_{0}}\right)
Y'_{P(z)}(v, x_{1})\lambda  \;\;\;\;\;
\mbox{\rm for all}\;\;v\in V.\label{comp}
\end{eqnarray}
Also, we say that $W$ is ($\Bbb{C}$-){\it graded} if it is
$\Bbb{C}$-graded by its weight subspaces, and that $W$ is a $V$-{\it module}
(respectively, {\it generalized module}) if $W$ is graded and is a
module (respectively, generalized module, see \cite{HL1} and
\cite{HL2}) when equipped with this grading and with the action of
$Y'_{P(z)}(\cdot, x)$. The weight subspace of a subspace $W$ with
weight $n\in \Bbb{C}$ will be denoted $W_{(n)}$.

Define
\begin{equation}
W_{1}\hboxtr_{P(z)}W_{2}=\sum_{W\in \cal{W}_{P(z)}}W =\bigcup_{W\in
\cal{W}_{P(z)}} W\subset
(W_{1}\otimes W_{2})^{*},
\end{equation}
where $\cal{W}_{P(z)}$ is the set of all compatible
modules for $\tau_{P(z)}$ in $(W_{1}\otimes W_{2})^{*}$.
We have:
\begin{propo}\label{rational}
Let $V$ be a rational vertex operator algebra and $W_{1}$, $W_{2}$
$V$-modules. Then $(W_{1}\hboxtr _{P(z)}W_{2},
Y'_{P(z)}\text{\huge $\vert$}_{V\otimes W_{1}\shboxtr _{P(z)}W_{2}})$
is a module.
\end{propo}

Now we assume that $V$ is rational.  In this case, we define a
$V$-module $W_{1}\boxtimes_{P(z)} W_{2}$ by
\begin{equation}
W_{1}\boxtimes_{P(z)} W_{2}=(W_{1}\hboxtr_{P(z)}W_{2})'
\end{equation}
 and we write the corresponding action as $Y_{P(z)}$. Applying
Proposition \ref{intwact} to the special module
$W_{3}=W_{1}\boxtimes_{P(z)} W_{2}$ and the identity map $W'_{3}\to
W_{1}\hboxtr_{P(z)} W_{2}$, we obtain a canonical
$P(z)$-intertwining map of type ${W_{1}\boxtimes_{P(z)} W_{2}}\choose
{W_{1}W_{2}}$, which we denote
\begin{eqnarray}
\boxtimes_{P(z)}: W_{1}\otimes W_{2}&\to &
\overline{W_{1}\boxtimes_{P(z)} W_{2}}\nno\\
w_{(1)}\otimes  w_{(2)}&\mapsto& w_{(1)} \boxtimes_{P(z)}w_{(2)}.
\end{eqnarray}
We have:

\begin{propo}
The $P(z)$-product $(W_{1}\boxtimes_{P(z)} W_{2}, Y_{P(z)};
\boxtimes_{P(z)})$
is a $P(z)$-tensor product of $W_{1}$ and $W_{2}$.
\end{propo}

Observe that any element of
$W_{1}\hboxtr_{P(z)} W_{2}$ is an element $\lambda$
of $(W_{1}\otimes W_{2})^{*}$ satisfying:

\begin{description}
\item[The compatibility condition] \hfill

{\bf (a)} The  {\it lower
truncation condition}:
For all $v\in V$, the formal Laurent series $Y'_{P(z)}(v, x)\lambda$
involves only finitely many negative
powers of $x$.

{\bf (b)} The formula (\ref{comp}) holds.

\item[The local grading-restriction  condition]\hfill

{\bf (a)} The {\it grading condition}:
$\lambda$ is a (finite) sum of
weight vectors of $(W_{1}\otimes W_{2})^{*}$.

{\bf (b)} Let $W_{\lambda}$ be the smallest subspace of $(W_{1}\otimes
W_{2})^{*}$ containing $\lambda$ and stable under the component
operators $\tau_{P(z)}(v\otimes t^{n})$ of the operators $Y'_{P(z)}(v,
x)$ for $v\in V$, $n\in \Bbb{Z}$. Then the weight spaces
$(W_{\lambda})_{(n)}$, $n\in \Bbb{C}$, of the (graded) space
$W_{\lambda}$ have the properties
\begin{eqnarray}
&\mbox{\rm dim}\ (W_{\lambda})_{(n)}<\infty \;\;\;\mbox{\rm for}\
n\in \Bbb{C},&\\
&(W_{\lambda})_{(n)}=0 \;\;\;\mbox{\rm for $n$ whose real part is
sufficiently small.}&
\end{eqnarray}
\end{description}

We have
another construction
of $W_{1}\hboxtr_{P(z)} W_{2}$ using these conditions:

\begin{theo}
The subspace of $(W_{1}\otimes W_{2})^{*}$ consisting of the
elements satisfying the compatibility
condition and the local grading-restriction condition, equipped with
$Y'_{P(z)}$, is a generalized module and is equal to
$W_{1}\hboxtr_{P(z)} W_{2}$.
\end{theo}

The following result follows immediately {from} Proposition
\ref{rational}, the theorem above and the definition of
$W_{1}\boxtimes_{P(z)} W_{2}$:

\begin{corol}
Let $V$ be a rational vertex operator algebra and $W_{1}$, $W_{2}$ two
$V$-modules.  Then the contragredient module of the module $W_1
\hboxtr_{P(z)} W_2$,
equipped with the $P(z)$-intertwining map $\boxtimes_{P(z)}$,
 is a $P(z)$-tensor product of $W_{1}$ and
$W_{2}$  equal to the structure $(W_{1}\boxtimes_{P(z)}
W_{2}, Y_{P(z)};
\boxtimes_{P(z)})$ constructed above.
\end{corol}

\subsection{The associativity}

Given any $V$-modules $W_{1}$, $W_{2}$, $W_{3}$, $W_{4}$ and $W_{5}$,
let $\cal{Y}_{1}$, $\cal{Y}_{2}$, ${\cal Y}_{3}$ and ${\cal Y}_{4}$
be intertwining operators of type ${W_{4}}\choose {W_{1}W_{5}}$,
${W_{5}}\choose {W_{2}W_{3}}$, ${W_{5}}\choose {W_{1}W_{2}}$ and
${W_{4}}\choose {W_{5}W_{3}}$, respectively. Consider the following
conditions for the product of $\cal{Y}_{1}$ and $\cal{Y}_{2}$ and
for the iterate of $\cal{Y}_{3}$ and $\cal{Y}_{4}$, respectively:

\begin{description}

\item[Convergence and extension property for products]
There exists
an integer $N$
(depending only on $\cal{Y}_{1}$ and $\cal{Y}_{2}$), and
for any $w_{(1)}\in W_{1}$,
$w_{(2)}\in W_{2}$, $w_{(3)}\in W_{3}$, $w'_{(4)}\in W'_{4}$, there exist
$j\in \Bbb{N}$, $r_{i}, s_{i}\in \Bbb{R}$, $i=1, \dots, j$, and analytic
functions $f_{i}(z)$ on $|z|<1$, $i=1, \dots, j$,
satisfying
\begin{equation}
\Re(\wt w_{(1)}+\wt w_{(2)}+s_{i})>N,\;\;\;i=1, \dots, j,
\end{equation}
such that
\begin{equation}
\langle w'_{(4)}, \cal{Y}_{1}(w_{(1)}, x_{2})
\cal{Y}_{2}(w_{(2)}, x_{2})w_{(3)}\rangle_{W_{4}}
\lbar_{x_{1}^{n}= e^{n\log z_{1}}, \;x_{2}^{n}=e^{n\log z_{2}},\;
n\in \Bbb{C}}
\end{equation}
is convergent when $|z_{1}|>|z_{2}|>0$ and can be analytically extended to
the multi-valued analytic function
\begin{equation}\label{phyper}
\sum_{i=1}^{j}z_{2}^{r_{i}}(z_{1}-z_{2})^{s_{i}}
f_{i}\left(\frac{z_{1}-z_{2}}{z_{2}}\right)
\end{equation}
when $|z_{2}|>|z_{1}-z_{2}|>0$.

\item[Convergence and extension property for iterates]
\hspace{.5em}There exists an integer
$\tilde{N}$
(depending only on $\cal{Y}_{3}$ and $\cal{Y}_{4}$), and
for any $w_{(1)}\in W_{1}$,
$w_{(2)}\in W_{2}$, $w_{(3)}\in W_{3}$, $w'_{(4)}\in W'_{4}$, there exist
$k\in \Bbb{N}$, $\tilde{r}_{i}, \tilde{s}_{i}\in \Bbb{R}$, $i=1, \dots, k$,
and analytic
functions $\tilde{f}_{i}(z)$ on $|z|<1$, $i=1, \dots, k$,
satisfying
\begin{equation}
\Re(\wt w_{(2)}+\wt w_{(3)}+s_{i})>\tilde{N},\;\;\;i=1, \dots, k,
\end{equation}
 such that
\begin{equation}
\langle w'_{(4)},
\cal{Y}_{4}(\cal{Y}_{3}(w_{(1)}, x_{0})w_{(2)}, x_{2})w_{(3)}\rangle_{W_{4}}
\lbar_{x^{n}_{0}=e^{n\log (z_{1}-z_{2})},\;x^{n}_{2}=e^{n\log z_{2}},\;
n\in \Bbb{C}}
\end{equation}
is convergent when $|z_{2}|>|z_{1}-z_{2}|>0$ and can be analytically extended
to the multi-valued analytic function
\begin{equation}
\sum_{i=1}^{k}z_{1}^{\tilde{r}_{i}}z_{2}^{\tilde{s}_{i}}
\tilde{f}_{i}\left(\frac{z_{2}}{z_{1}}\right)
\end{equation}
when $|z_{1}|>|z_{2}|>0$.

\end{description}

If for any $V$-modules $W_{1}$, $W_{2}$, $W_{3}$, $W_{4}$ and $W_{5}$ and
 any intertwining operators $\cal{Y}_{1}$ and $\cal{Y}_{2}$
of the types as above, the convergence and extension property for
products holds,
we say that
{\it  the products of the
 intertwining operators for $V$ have the convergence and extension property}.
Similarly we can define what {\it the iterates of the intertwining
operators for $V$ have the  convergence and extension property}
means.

If a generalized $V$-module $W=\coprod_{n\in \Bbb{C}}W_{(n)}$
satisfying the condition that $W_{(n)}=0$ for $n$ whose real part is
sufficiently small, we say that $W$ is {\it lower-truncated}.

Assume that the products and the iterates of the intertwining
operators for $V$ are convergent. Let $W_{1}$, $W_{2}$ and $W_{3}$ be
three $V$-modules, $w_{(1)}\in W_{1}$, $w_{(2)}\in W_{2}$ and
$w_{(3)}\in W_{3}$ and $z_{1}, z_{2}\in \Bbb{C}$ satisfying
$|z_{1}|>|z_{2}|>|z_{1}-z_{2}|>0$. By Proposition \ref{op-map},
any $P(z)$-intertwining maps (for
$z=z_{1}, z_{2}, z_{1}-z_{2}$) can be obtained {}from
certain intertwining
operators by substituting complex powers of $e^{\log z}$ for the
complex powers of the formal variable $x$. Thus
$w_{(1)}\boxtimes_{P(z_{1})}(w_{(2)}\boxtimes_{P(z_{2})}w_{(3)})$ (or
$(w_{(1)}\boxtimes_{P(z_{1}-z_{2})}w_{(2)})\boxtimes_{P(z_{2})}w_{(3)}$)
is a product (or an iterate) of two intertwining operators evaluated
at $w_{(1)}\otimes w_{(2)}\otimes w_{(3)}$ and with the complex powers
of the formal variables replaced by the complex powers of $e^{\log
z_{1}}$ and of $e^{\log z_{2}}$ (or by the complex powers of $e^{\log
(z_{1}-z_{2})}$ and of $e^{\log z_{2}}$). By assumption,
$w_{(1)}\boxtimes_{P(z_{1})}(w_{(2)}\boxtimes_{P(z_{2})}w_{(3)})$ (or
$(w_{(1)}\boxtimes_{P(z_{1}-z_{2})}w_{(2)})\boxtimes_{P(z_{2})}w_{(3)}$)
 is a well-defined element of
$\overline{W_{1}\boxtimes_{P(z_{1})}(W_{2}\boxtimes_{P(z_{2})}W_{3})}$
(or of $\overline{(W_{1}\boxtimes_{P(z_{1}-z_{2})}W_{2})
\boxtimes_{P(z_{2})}W_{3}}\;$).
The following result is proved in \cite{H2}:

\begin{theo}\label{assoc}
Assume that $V$ is a rational vertex operator algebra and
all irreducible $V$-modules are $\Bbb{R}$-graded.  Also assume
that $V$ satisfies the following conditions:
\begin{enumerate}
\item Every finitely-generated lower-truncated generalized $V$-module
is a $V$-module.

\item The products or the iterates of
the intertwining operators for $V$ have the  convergence and
extension property.
\end{enumerate}
Then for any $V$-modules $W_{1}$, $W_{2}$ and $W_{3}$ and any complex
numbers $z_{1}$ and $z_{2}$ satisfying
$|z_{1}|>|z_{2}|>|z_{1}-z_{2}|>0$, there exists a unique isomorphism
$\cal{A}_{P(z_{1}), P(z_{2})}^{P(z_{1}-z_{2}), P(z_{2})}$ {}from
$W_1\boxtimes_{P(z_1)}(W_2 \boxtimes_{P(z_2)}W_3)$ to $(W_1
\boxtimes_{P(z_1-z_2)}W_2)\boxtimes_{P(z_2)}W_3$ such that
for any $w_{(1)}\in W_{1}$, $w_{(2)}\in W_{2}$ and $w_{(3)}\in W_{3}$,
\begin{eqnarray*}
\lefteqn{\overline{\cal{A}}_{P(z_{1}), P(z_{2})}^{P(z_{1}-z_{2}),
P(z_{2})}(w_{(1)}\boxtimes_{P(z_{1})}(w_{(2)}
\boxtimes_{P(z_{2})}w_{(3)}))}\nno\\
&&=(w_{(1)}\boxtimes_{P(z_{1}-z_{2})}w_{(2)})
\boxtimes_{P(z_{2})}w_{(3)},
\end{eqnarray*}
where $$\overline{\cal{A}}_{P(z_{1}), P(z_{2})}^{P(z_{1}-z_{2}),
P(z_{2})}: \overline{W_1\boxtimes_{P(z_1)}(W_2 \boxtimes_{P(z_2)}W_3)}
\to \overline{(W_1 \boxtimes_{P(z_1-z_2)}W_2)\boxtimes_{P(z_2)}W_3}$$
is the unique extension of $\cal{A}_{P(z_{1}),
P(z_{2})}^{P(z_{1}-z_{2}), P(z_{2})}$.
\end{theo}

\section{The nonmeromorphic extension of $V^{\natural}$}

We sketch the construction the nonmeromorphic extension of
$V^{\natural}$ in this section.  This nonmeromorphic extension is in
fact the algebra of all intertwining operators for the vertex operator
algebra $V_{\Lambda}^{+}$.  We first verify that the conditions for
the tensor product theory reviewed in Section 2 are satisfied by
$V_{\Lambda}^{+}$. Then we calculate the fusion rules for
$V_{\Lambda}^{+}$. The nonmeromorphic extension is obtained using the
fusion rules and the tensor product theory. The details of the proofs of the
results stated in Subsections 3.1 and 3.2 are given in \cite{H3}.

\subsection{Modules for the Virasoro vertex operator algebras and
their tensor products} To
prove that $V_{\Lambda}^{+}$ satisfies the conditions in Theorem
\ref{assoc}, we shall use the results in \cite{DMZ} and \cite{D3}.  So
we first have to prove that the tensor product theory can be applied
to the vertex operator algebra $L(\frac{1}{2}, 0)$.  The rationality
of $L(\frac{1}{2}, 0)$ is proved in \cite{DMZ}. In general, the
rationality of the Virasoro vertex operator algebra $L(c_{p, q}, 0)$ of
central
charge $c_{p, q}=1-6\frac{(p-q)^{2}}{pq}$ is proved in \cite{W} for an
arbitrary pair $p, q$ of relatively prime positive integers larger
than $1$.
We have the following result for $L(c_{p, q}, 0)$:

\begin{propo}\label{cpq}
Let $p, q$ be a pair of relatively prime positive
integers larger than $1$. Then we have:
\begin{enumerate}

\item Every finitely-generated lower-truncated generalized
$L(c_{p, q}, 0)$-module is a module.

\item The products of the intertwining operators for $L(c_{p, q}, 0)$
 have the convergence and extension property.

\end{enumerate}
\end{propo}
{\it Sketch of the proof}\hspace{2ex}
The first conclusion is an easy exercise on
 the representations of the Virasoro algebra.
The second conclusion is proved using the representation theory of
the Virasoro algebra and the differential equations of
Belavin, Polyakov and Zamolodchikov (BPZ equations)
for correlation functions in the
minimal models \cite{BPZ}.  We show that
\begin{equation}
\langle w'_{(4)}, \cal{Y}_{1}(w_{(1)}, x_{1})
\cal{Y}_{2}(w_{(2)}, x_{2})w_{(3)}\rangle\lbar_{x_{1}^{n}=e^{n\log
z_{1}}, x_{2}^{n}=e^{n \log z_{2}}, n\in \Bbb{C}}\label{correl}
\end{equation}
 satisfies a system of BPZ equations when $w_{(1)}, w_{(2)}, w_{(3)}$ and
$w'_{(4)}$ are all lowest weight vectors.  The BPZ equation has only
regular singular points. Thus using the theory of equations of regular
singular points (see, for example, Appendix B of \cite{K})
and the definition of intertwining operator,
we can show that (\ref{correl}) in this case is convergent when
$|z_{1}|>|z_{2}|>0$ and can be analytically extended to a function
of the form (\ref{phyper}) when $|z_{2}|>|z_{1}-z_{2}|>0$. This together
with the brackets of $L(n)$, $n \in \Bbb{C}$, with the intertwining
operators and the $L(-1)$-derivative property for the intertwining
operators shows that  the products of the intertwining operators for
$L(c_{p, q}, 0)$ have the  convergence and extension property.
\epfv

Next we discuss tensor products of $L(c_{p, q}, 0)$.

\begin{lemma}\label{intwop}
Let $n$ be a positive integer, $(p_{i}, q_{i})$, $i=1, \dots, n$, $n$
pairs of relatively prime positive integers larger than $1$,
$V=L(c_{p_{1}, q_{1}}, 0)\otimes \cdots \otimes L(c_{p_{n}, q_{n}},
0)$, $W_{i}=L(c_{p_{1}, q_{1}}, h_{1}^{(i)})\otimes \cdots
\otimes L(c_{p_{n}, q_{n}}, h_{n}^{(i)})$,
$i=1, 2, 3$, irreducible $V$-modules and $\cal{Y}$ an
intertwining operator of type ${W_{3}}\choose {W_{1}W_{2}}$. Then
there exist intertwining operators $\cal{Y}_{i}$
of type ${L(c_{p_{i}, q_{i}}, h_{i}^{(3)})} \choose
{L(c_{p_{i}, q_{i}}, h_{i}^{(1)})L(c_{p_{i}, q_{i}}, h_{i}^{(2)})}$,
$i=1, \dots, n$, such that
\begin{eqnarray*}
\cal{Y}=\cal{Y}_{1}\otimes \cdots \otimes \cal{Y}_{n}.
\end{eqnarray*}
\end{lemma}

This lemma is an easy consequence of the result in \cite{DMZ}
expressing the fusion rules of $V$ in terms of the fusion rules of
$L(c_{p_{i}, q_{i}}, 0)$, $i=1, \dots, n$. It can also be proved directly
using the special properties of the Virasoro vertex operator algebras.
 Using Proposition
\ref{cpq}, Lemma \ref{intwop} and the methods used to prove results on
modules for tensor products of a vertex operator algebra in
\cite{FHL}, we obtain:

\begin{propo}\label{cpqn}
For any positive integer $n$ and any
$n$ pairs $(p_{i}, q_{i})$ of relatively prime positive
integers larger than $1$, $i=1, \dots, n$, we have:

\begin{enumerate}

\item Every finitely-generated lower-truncated generalized
$L(c_{p_{1}, q_{1}}, 0)\otimes \cdots \otimes
L(c_{p_{n}, q_{n}}, 0)$-module is a module.

\item The products of the intertwining operators for
$$L(c_{p_{1}, q_{1}}, 0)\otimes \cdots \otimes
L(c_{p_{n}, q_{n}}, 0)$$ have the convergence and extension property.

\end{enumerate}
\end{propo}

\subsection{A class of vertex operator algebras and the
associativity of tensor products} Let $n$ be a positive integer,
$(p_{i}, q_{i})$, $i=1, \dots, n$, $n$ pairs of relatively prime
positive integers larger than $1$.  A vertex operator algebra $V$ is
said to be in the class $\cal{C}_{p_{1}, q_{1};\dots; p_{n}, q_{n}}$
if $V$ has a vertex operator subalgebra isomorphic to $L(c_{p_{1}, q_{1}},
0)\otimes \cdots \otimes L(c_{p_{n}, q_{n}}, 0)$. The work of Dong,
Mason and Zhu \cite{DMZ} shows that $V^{+}_{\Lambda}$ is in the class
$\cal{C}_{3, 4; \dots; 3, 4}$ with $n=48$.

Using Proposition \ref{cpqn}, we can prove:

\begin{propo}\label{class}
Let $n$ be a positive integer,
$(p_{i}, q_{i})$, $i=1, \dots, n$, $n$ pairs of relatively prime
positive integers larger than $1$, and
let $V$ be a vertex operator algebra in the class $\cal{C}_{p_{1},
q_{1};
\dots; p_{n}, q_{n}}$.
Then we have:

\begin{enumerate}

\item Every finitely-generated lower-truncated generalized
$V$-module is a module.

\item The products of the intertwining operators for
$V$  have the convergence and extension property.

\end{enumerate}
\end{propo}

The proof of the second conclusion is easy.  The proof of the first
conclusion is more subtle than what it seems to be since a
finitely-generated generalized $V$-module is not obviously
finitely-generated as a generalized $L(c_{p_{1}, q_{1}}, 0)\otimes
\cdots \otimes L(c_{p_{n}, q_{n}}, 0)$-module.

\subsection{The fusion rules for $V_{\Lambda}^{+}$} We first
quote a result proved in \cite{FHL} and \cite{HL3}:

\begin{propo}\label{s3}
Let $V$ be a vertex operator algebra and $W_{1}$, $W_{2}$, $W_{3}$
three $V$-modules. Then for any permutation $\sigma$ of three letters,
$\cal{N}_{W_{\sigma(1)}W_{\sigma(2)}}^{W'_{\sigma(3)}} ={\cal
N}_{W_{1}W_{2}}^{W'_{3}}$. In particular, if $W_{1}$, $W_{2}$ and
$W_{3}$ are all self-dual, that is, are isomorphic to their
contragredient modules $W'_{1}$, $W'_{2}$ and $W'_{3}$, respectively,
then $\cal{N}_{W_{\sigma(1)}W_{\sigma(2)}}^{W_{\sigma(3)}} ={\cal
N}_{W_{1}W_{2}}^{W_{3}}$.
\end{propo}

In the special case that one of the three modules is the adjoint
module and the other two are irreducible, it is easy to prove:

\begin{propo}\label{vww}
Let $V$ be a vertex operator algebra and $W_{1}$, $W_{2}$ irreducible
$V$-modules contragredient to themselves.
Then the fusion rules $\cal{N}_{W_{1}W_{2}}^{V}$, ${\cal
N}_{VW_{1}}^{W_{2}}$ and $\cal{N}_{W_{1}V}^{W_{2}}$ are equal to $0$
if $W_{1}$ and $W_{2}$ are not isomorphic and are equal to $1$ if they
are isomorphic.
\end{propo}

For $i, j\in \Bbb{Z}_{2}$, let
$$M^{i, j}=\left\{\begin{array}{ll}V_{\Lambda}^{+},&i=j=0,\\
V_{\Lambda}^{-},&i=1, j=0,\\ (V_{\Lambda}^{T})^{+},&i=0, j=1,\\
(V_{\Lambda}^{T})^{-},&i=j=1.\end{array}\right.$$

\begin{theo}\label{fusion}
The fusion rules for $V_{\Lambda}^{+}$ is
$$\cal{N}_{M^{i_{1}, j_{1}}M^{i_{2}, j_{2}}}^{M^{i_{3}, j_{3}}}
=\left\{\begin{array}{ll}1, &i_{3}=i_{1}+i_{2}, j_{3}=j_{1}+j_{2},\\
0,& \mbox{\rm otherwise}. \end{array}\right.$$
\end{theo}
{\it Sketch of the proof}\hspace{2ex} The proof uses Proposition
\ref{s3}, Proposition~\ref{vww}, Theorem~\ref{dmz},
Theorem~\ref{decomp}, Theorem~\ref{d3} and the tensor product theory
described in Section 2, especially the associativity for
$P(\cdot)$-tensor products. Using the first half of Theorem \ref{d3},
we can prove $\cal{N}_{M^{i_{1}, 1}M^{i_{2}, 1}}^{M^{i_{3}, 1}}=0$
and $\cal{N}_{M^{i_{1}, 0}M^{i_{2}, 0}}^{M^{i_{3}, 1}}=0$.  These
fusion rules together with Proposition
\ref{s3} and Proposition \ref{vww} reduces the problem to the
calculations of ${\cal
N}^{(V_{\Lambda}^{T})^{-}}_{V_{\Lambda}^{-}(V_{\Lambda}^{T})^{+}}$,
${\cal
N}^{(V_{\Lambda}^{T})^{-}}_{V_{\Lambda}^{-}(V_{\Lambda}^{T})^{-}}$,
$\cal{N}^{(V_{\Lambda}^{T})^{+}}_{V_{\Lambda}^{-}
(V_{\Lambda}^{T})^{+}}$ and
$\cal{N}^{V_{\Lambda}^{-}}_{V_{\Lambda}^{-} V_{\Lambda}^{-}}$. The
calculations of these fusion rules use more details on the lowest
weight vectors with respect to $L(\frac{1}{2}, 0)^{\otimes 48}$ in
$V_{\Lambda}^{\pm}$ and $(V_{\Lambda}^{T})^{\pm}$ given in
Theorem~\ref{decomp} and the associativity of the intertwining
operators which is in fact equivalent to the associativity of the
$P(\cdot)$-tensor products. By Theorem \ref{do1}, we know that
 $V_{\Lambda}^{+}$ is rational and the four irreducible
$V_{\Lambda}^{+}$-modules are all $\Bbb{R}$-graded.  The rationality of
$V_{\Lambda}^{+}$, the fact that all $V_{\Lambda}^{+}$-modules
are $\Bbb{R}$-graded and Proposition \ref{class} guarantee that we can
use the tensor product theory, especially the associativity.
\epf

\subsection{The nonmeromorphic extension} Let
$$W^{\natural}=V_{\Lambda}\oplus V_{\Lambda}^{+}=\coprod_{(i, j)\in
\Bbb{Z}_{2}\oplus
\Bbb{Z}_{2}}M^{i, j}.$$ We define
a vertex operator map $Y_{W^{\natural}}: W^{\natural}
\otimes W^{\natural}\to W^{\natural}[[x^{\frac{1}{2}},
x^{-\frac{1}{2}}]]$ as follows: For $u, v\in V_{\Lambda}$,
$Y_{W^{\natural}}(u, x)v=Y_{V_{\Lambda}}(u, v)$; for $u\in
V_{\Lambda}$, $v\in V_{\Lambda}^{T}$, $Y_{W^{\natural}}(u,
x)v=Y_{V_{\Lambda}^{T}}(u, x)v$; for $u\in V_{\Lambda}^{T}$, $v\in
V_{\Lambda}$, $Y_{W^{\natural}}(u,
x)v=e^{xL(-1)}Y_{V_{\Lambda}^{T}}(v, e^{\pi i}x)u$; for $u, v\in
V_{\Lambda}^{T}$, $Y_{W^{\natural}}(u, x)v$ is defined by $$\langle w,
Y_{W^{\natural}}(u, x)v\rangle_{V_{\Lambda}} =\langle
Y_{V_{\Lambda}^{T}}(w, e^{\pi i}x^{-1})e^{xL(1)}(e^{\pi
i}x^{2})^{-L(0)}u, e^{x^{-1}L(1)}v\rangle_{V_{\Lambda}^{T}}$$
for all $w\in V_{\Lambda}$. We have
a (nonsymmetric) nondegenerate $\{1, -1\}$-valued $\Bbb{Z}$-bilinear
form $\Omega_{SU}$ (the subscript $SU$ means super, see
Subsection 4.2) on the finite abelian group $\Bbb{Z}_{2}\oplus
\Bbb{Z}_{2}$ determined uniquely by $\Omega_{SU}((1, 0), (1, 0))=1$,
$\Omega_{SU}((1, 0), (0, 1))=-1$, $\Omega_{SU}((0, 1), (1, 0))=1$,
$\Omega_{SU}((0, 1), (0, 1))=1$, and the bilinearity. To formulate the
main result of this paper, we need the notions of
 abelian intertwining algebra, whose definition,
examples and axiomatic properties can be found in \cite{DL}.
In the definition of abelian intertwining algebra, part of
the data is an abelian group $G$
and a normalized abelian 3-cocycle $(F, \Omega)$ for the abelian group
$G$ with values in $\Bbb{C}^{\times}$,
where $F$ is a normalized 3-cocycle for $G$ as a group.
In this paper,  $G$ is $\Bbb{Z}_{2}\oplus \Bbb{Z}_{2}$,
$F$ is trivial (denoted by $1$) and $\Omega$ is equal to $\Omega_{SU}$
defined above.

\begin{theo}\label{main}
The structure $(W^{\natural}, Y_{W^{\natural}}, \bold{1}, \omega, 2,
\Bbb{Z}_{2}\oplus \Bbb{Z}_{2}, 1, \Omega_{SU})$ is an
abelian intertwining algebra of central charge $24$.
\end{theo}
{\it Sketch of the proof}\hspace{2ex} The proof uses the fusion rules
in Theorem \ref{fusion} and the tensor product theory described in
Section 2, especially the associativity. We already know that the
tensor product theory can be applied to $V_{\Lambda}^{+}$.  {}From \cite{H2},
 we know that the
associativity of $P(\cdot)$-tensor products is equivalent to the
associativity of the intertwining operators.  So in this case, we
have the associativity for intertwining operators.  By the fusion rules,
for every ordered triple of irreducible $V_{\Lambda}^{+}$-modules,
any two intertwining operators of the type specified by this triple are
linearly dependent. On the other hand,
$Y_{W^{\natural}}$ restricted to any ordered triple of
irreducible $V_{\Lambda}^{+}$-modules
is an intertwining operator and is nonzero if the fusion rule is nonzero.
Thus the associativity for intertwining operators gives
the associativity for $Y_{W^{\natural}}$. It can be verified directly
that $Y_{W^{\natural}}$ satisfies a version of the skew symmetry for vertex
operators. Combining the associativity and the skew symmetry of
$Y_{W^{\natural}}$, we obtain the commutativity of $Y_{W^{\natural}}$.
It can be shown that if the fusion algebra for a rational
vertex operator algebra satisfying the conditions in Theorem \ref{assoc}
is the group algebra of an abelian group, the products and the
iterates of intertwining operators among irreducible modules
 must be appropriate expansions of
generalized rational functions (see \cite{DL} for the meaning of
generalized rational functions).
In our case, the fusion algebra is the group algebra of the abelian group
$\Bbb{Z}_{2}\oplus \Bbb{Z}_{2}$. Thus we have the generalized
rationalities of both products and iterates for $Y_{W^{\natural}}$.
\epf

\section{Applications}

We give applications of the results obtained in the preceding section.
A special case of Theorem \ref{main} gives a new and conceptual proof
that $V^{\natural}$ has a natural vertex operator algebra structure.
Other special cases give an irreducible
twisted module for the moonshine module with respect to the obvious
involution, a vertex operator superalgebra and a twisted module for
this vertex operator superalgebra with respect to the involution
which is the identity on the even subspace and is $-1$ on
the odd subspace.
  We also use Theorem \ref{main} to construct
the superconformal structures of Dixon, Ginsparg and Harvey
rigorously.

\subsection{The moonshine module and its twisted module}
When we restrict ourselves to the moonshine module
$V^{\natural}=V_{\Lambda}^{+}\oplus (V_{\Lambda}^{T})^{+}\subset
W^{\natural}$, we immediately obtain the following:

\begin{theo}\label{moonshine}
The quadruple $(V^{\natural}, Y_{V^{\natural}}, \bold{1}, \omega)$ is a
vertex operator algebra.
\end{theo}

\begin{rema}
Unlike the proof of this theorem given in \cite{FLM2}, our proof does
not use triality or any result in group theory.  It can be proved
without using triality or group theory that any automorphism of
the Griess algebra can be extended to an automorphism of the vertex
operator algebra $V^{\natural}$ (this was also observed by Dong).  Thus
our proof (or any proof without using  triality or group theory,
for example, the one given in \cite{DGM1}) makes the fact that the
automorphism group of the Griess algebra and the automorphism group of
the vertex operator algebra $V^{\natural}$ are isomorphic, to be
logically independent of triality and group theory. This
independence allows us to obtain another proof of the theorem saying
that the Monster is the (full) automorphism group of the vertex
operator algebra $V^{\natural}$ based on the the theorem saying that
the Monster is the (full) automorphism group of the Griess algebra
proved by Griess \cite{G} and Tits \cite{Ti1} \cite{Ti2}, simplified
by Conway \cite{C} and Tits \cite{Ti2} and understood conceptually by
Frenkel, Lepowsky and Meurman using the theory of vertex operators
and triality \cite{FLM1}
\cite{FLM2}.
\end{rema}

Another immediate consequence is on the irreducible
twisted module for $V^{\natural}$:

\begin{theo}\label{twistedmdle}
Let $\tau: V^{\natural}\to V^{\natural}$ be an involution defined by
$\tau(v)=v$ if $v\in V_{\Lambda}^{+}$ and $\tau(v)=-v$ if $v\in
(V_{\Lambda}^{T})^{+}$.  Then the pair $$(V_{\Lambda}^{-}\oplus
(V_{\Lambda}^{T})^{-}, Y_{W^{\natural}}
\text{\huge $\vert$}_{V^{\natural}\otimes (V_{\Lambda}^{-}\oplus
(V_{\Lambda}^{T})^{-})})$$ is an irreducible $\tau$-twisted module for
$V^{\natural}$. Any irreducible $\tau$-twisted module is isomorphic to
this one.
\end{theo}

Theorem \ref{twistedmdle}
is also proved by Dong and Mason using a different method.

\begin{rema}
Note that if we are only interested in Theorem \ref{moonshine} or
Theorem \ref{twistedmdle}, we can prove them using only parts of the
fusion rules which we calculated for $V_{\Lambda}^{+}$.
\end{rema}

\subsection{The superconformal structures of Dixon,
Ginsparg and Harvey}
The following result  observed first by Dixon,
Ginsparg and Harvey is proved using Theorem \ref{main}
and some concrete calculations of twisted vertex operators:

\begin{theo}\label{super}
For any $t\in T$ satisfying $\langle 1\otimes t, 1\otimes
t\rangle_{V^{T}_{\Lambda}}=1$ (for example, $t(a)$ for any $a\in
\hat{\Lambda}$), let $Y_{W^{\natural}}(2(1\otimes t), x)
=\sum_{n\in
\frac{1}{2}\Bbb{Z}} G(n)x^{-n-\frac{3}{2}}$. Then the operators
$L(n)$, $n\in \Bbb{Z}$ and $G(n)$, $n\in \frac{1}{2}\Bbb{Z}$,
satisfies the super-Virasoro relations:
\begin{eqnarray}
[L(m), L(n)]&=&(m-n)L(m+n)+\frac{\hat{c}_{W^{\natural}}}
{8}(m^{3}-m)\delta_{m+n, 0},
\label{svir1}\\
{[L(m), G(n)]}&=&\left(\frac{m}{2}-n\right)G(m+n),\label{svir2}\\
\{G(m), G(n)\}&=&2L(m+n)+\frac{\hat{c}_{W^{\natural}}}{2}
\left(m^{2}-\frac{1}{4}\right)\delta_{m+n, 0}\label{svir3}
\end{eqnarray}
where the super-central charge $\hat{c}_{W^{\natural}}=16$ and
$\{\cdot, \cdot\}$ denotes the anti-bracket.
\end{theo}

To summarize the superconformal structures on $W^{\natural}$,
we need the following notions:

\begin{defi}
An {\it (N=1) Neveu-Schwarz type superconformal vertex operator algebra
of super-central charge} (or {\it super-rank}) $\hat{c}$
is a vertex operator superalgebra $(V, Y, \bold{1}, \omega)$
(see, for example,
\cite{FFR} or \cite{DL}), equipped with an element $\xi\in V$ such that the
components of $Y(x^{L(0)}\omega, x)$ and $Y(x^{L(0)}\xi, x)$ satisfies the
super-Virasoro relations (\ref{svir1})--(\ref{svir3}) with
$\hat{c}_{W^{\natural}}$ replaced by $\hat{c}$.
\end{defi}

In \cite{KW}, (N=1) Neveu-Schwarz type superconformal vertex operator algebras
are  studied and are called ``N=1
(NS-type) vertex operator superalgebras.''
The (N=1) Neveu-Schwarz type superconformal vertex operator algebra
just defined is denoted by $(V, Y, \bold{1}, \omega, \xi)$ or simply by $V$.
{\it Homomorphisms, isomorphisms and automorphisms} of (N=1)
Neveu-Schwarz type
superconformal vertex operator
algebras are defined in the obvious way.

Let $(V, Y, \bold{1}, \omega, \xi)$ be
a Neveu-Schwarz type superconformal vertex operator algebra of
super-central charge $\hat{c}$. A module for $V$ is defined to be
a module for $V$ as a vertex operator superalgebra.
We define a linear isomorphism
$\sigma$ of $V$ by linearity and
\begin{equation}
\sigma(v)=\left\{\begin{array}{cc} v,&v\in V^{0},\\
-v,&v\in V^{1}, \end{array}\right.
\end{equation}
where $V^{0}$ and $V^{1}$ are even and odd subspaces of $V$, respectively.
It is clear that $\sigma$ is an involution and an automorphism of
the Neveu-Schwarz type superconformal vertex operator algebra $V$.
Therefore it is natural to consider $\sigma$-twisted $V$-modules.
See \cite{FFR} for a definition of $\sigma$-twisted $V$-modules.
Following physicists'
terminology, we also
call an (untwisted) module for $V$ a {\it Neveu-Schwarz sector for $V$} and
a $\sigma$-twisted module for $V$ a {\it Ramond sector for $V$}.

\begin{defi}\label{sc}
An {\it (N=1) superconformal vertex operator algebra of super-central
charge} (or {\it super-rank}) $\hat{c}$ is a abelian intertwining
algebra $$(W, Y_{W}, \bold{1}, \omega, 2, \Bbb{Z}_{2}\oplus \Bbb{Z}_{2}, 1,
\Omega_{SU})$$
 (where $\Omega_{SU}$ is the $\Bbb{Z}$-bilinear
 form on $\Bbb{Z}_{2}\oplus \Bbb{Z}_{2}$ defined in
Subsection 3.4), equipped with an element $\xi\in W$ such that the
components of $Y(\omega, x)$ and $Y(\xi, x)$ satisfies the
super-Virasoro relations (\ref{svir1})--(\ref{svir3}) with
$\hat{c}_{W^{\natural}}$ replaced by $\hat{c}$. The element $\xi$ is
called the {\it Neveu-Schwarz-Ramond element}. Let $W=\coprod_{(i,
j)\in \Bbb{Z}_{2}
\oplus \Bbb{Z}_{2}}W^{i, j}$. Then $W_{NS}=
W^{0,0}\oplus W^{1, 1}$ is called the
{\it Neveu-Schwarz sector} and $W_{R}=
W^{0, 1}\oplus W^{1,0}$ is called the {\it
Ramond sector}.
\end{defi}

The abelian intertwining algebra underlying a superconformal vertex operator
algebra can be described using its substructures as follows:

\begin{propo}
Let $(W, Y_{W}, \bold{1}, \omega, 2, \Bbb{Z}_{2}\oplus \Bbb{Z}_{2}, 1,
\Omega_{SU})$ be an abelian intertwining algebra of central
charge $\frac{3}{2}\hat{c}$. Then we have:

\begin{enumerate}

\item\label{alg} The ${\Bbb Z}$-graded vector spaces
$W^{0, 0}$, $W^{0, 0}\oplus W^{0, 1}$,
$W^{0, 0}\oplus W^{1, 0}$ with the restrictions of $Y_{W}$
as the vertex operator maps, the vacuum $\bold{1}$
and the Virasoro element $\omega$ are vertex operator algebras of
central charge $\frac{3}{2}\hat{c}$ and $W_{NS}=
W^{0,0}\oplus W^{1, 1}$ is a
Neveu-Schwarz type superconformal vertex operator algebra of
super-central charge $\hat{c}$.

\item\label{mods} The $\frac{1}{2}{\Bbb Z}$-graded vector
spaces $W^{i, j}$, $i,j\in {\Bbb Z}_{2}$, are modules for
$W^{0, 0}$.

\item\label{twmods} The $\frac{1}{2}{\Bbb Z}$-graded vector
spaces $W^{1, 0}\oplus W^{1, 1}$ and
$W^{0, 1}\oplus W^{1, 1}$ with the restrictions of $Y_{W}$ as
the  vertex operator maps are
twisted modules for $W^{0, 0}\oplus W^{0, 1}$ and
$W^{0, 0}\oplus W^{1, 0}$, respectively, with respect to the
obvious involutions.

\item\label{stwmod} The $\frac{1}{2}{\Bbb Z}$-graded vector
spaces $W_{R}=
W^{0, 1}\oplus W^{1,0}$ with with the restrictions of $Y_{W}$ as
the  vertex operator maps
is a $\sigma$-twisted module for
$W_{NS}$.

\end{enumerate}

Conversely, let $W=\coprod_{i, j\in {\Bbb Z}_{2}}W^{i, j}$
where $W^{i, j}$, $i, j\in {\Bbb Z}_{2}$, are four
$\frac{1}{2}{\Bbb Z}$-graded vector spaces be equipped with a vertex
operator map
$Y_{W}: W\otimes W\to W\{x\}$ and two distinguished elements
$\bold{1}$ and $\omega$ such that
$(${\normalshape i}$)$--$(${\normalshape iv}$)$
above hold.
Then $$(W, Y_{W}, \bold{1}, \omega, 2, \Bbb{Z}_{2}\oplus \Bbb{Z}_{2}, 1,
\Omega_{SU})$$ is an abelian intertwining algebra.
\end{propo}

We can summarize the main result and the main applications of the present
paper as follows:

\begin{theo}
For any $t\in T$ satisfying $\langle 1\otimes t, 1\otimes
t\rangle_{V_{\Lambda}^{T}}=1$ (for example, $t(a)$ for any $a\in
\hat{\Lambda}$), $(W^{\natural}, Y_{W^{\natural}}, \bold{1},
\omega, 2,
\Bbb{Z}_{2}\oplus \Bbb{Z}_{2}, \Omega_{SU}, 2(1\otimes t))$
is a superconformal vertex operator algebra of super-central charge
$\hat{c}_{W^{\natural}}=16$. In particular, we have:

\begin{enumerate}

\item The moonshine module
$V^{\natural}=V_{\Lambda}^{+}\oplus (V_{\Lambda}^{T})^{+}$
with the restriction of $Y_{W^{\natural}}$
as the vertex operator map, the vacuum $\bold{1}$
and the Virasoro element $\omega$ is a vertex operator algebras of
central charge $\frac{3}{2}\hat{c}_{W^{\natural}}=24$ and
$W^{\natural}_{NS}=V_{\Lambda}^{+}\oplus (V_{\Lambda}^{T})^{-}$
with the restriction of $Y_{W^{\natural}}$
as the vertex operator map
is a
Neveu-Schwarz type superconformal vertex operator algebra of
super-central charge $\hat{c}_{W^{\natural}}=16$.

\item The $\frac{1}{2}{\Bbb Z}$-graded vector
spaces $V_{\Lambda}^{-}\oplus (V_{\Lambda}^{T})^{-}$
with the restriction of $Y_{W^{\natural}}$
as the vertex operator map is a
$\tau$-twisted modules for $V^{\natural}$.

\item The $\frac{1}{2}{\Bbb Z}$-graded vector
spaces $W^{\natural}_{R}=V_{\Lambda}^{-}\oplus (V_{\Lambda}^{T})^{+}$
with the restriction of $Y_{W^{\natural}}$
as the vertex operator map
is a $\sigma$-twisted module for $W^{\natural}_{NS}$.

\end{enumerate}
\end{theo}

\end{document}